\documentclass[prodmode,acmtkdd, acmtog]{acmsmall}

\usepackage{algorithm}
\usepackage{algorithmic}
\usepackage{subfigure}
\usepackage{multirow}

\acmVolume{}
\acmNumber{}
\acmArticle{}
\acmYear{}
\acmMonth{0}

\begin{document}

\markboth{M. Das et al.}{Top-K Product Design Based on Collaborative Tagging Data}

\title{Top-K Product Design Based on Collaborative Tagging Data}
\author{Mahashweta Das
\affil{University of Texas at Arlington}
Gautam Das
\affil{University of Texas at Arlington, Qatar Computing Research Institute}
Vagelis Hristidis
\affil{University of California Riverside}
}

\begin{abstract}
The widespread use and popularity of collaborative content sites (e.g., IMDB, Amazon, Yelp, etc.) has created rich resources for users to consult in order to make purchasing decisions on various products such as movies, e-commerce products, restaurants, etc. Products with desirable tags (e.g., {\tt modern}, {\tt reliable}, etc.) have higher chances of being selected by prospective customers. This creates an opportunity for product designers to design better products that are likely to attract desirable tags when published. In this paper, we investigate how to mine collaborative tagging data to decide the attribute values of new products and to return the top-$k$ products that are likely to attract the maximum number of desirable tags when published. Of course, real-world product design is a complex task, and tag desirability is only one - albeit novel - aspect of the design considerations. The motivation is that the returned set of $k$ products can assist product designers who can then select from among them using additional constraints such as price, profitability, etc. Given a training set of existing products with their features and user-submitted tags, we first build a Naive Bayes Classifier for each tag. We show that the problem of is NP-complete even if simple Naive Bayes Classifiers are used for tag prediction. We present a suite of algorithms for solving this problem: (a) an exact {\em two-tier} algorithm (based on top-$k$ querying techniques), which performs much better than the naive brute-force algorithm and works well for moderate problem instances, and (b) a set of approximation algorithms for larger problem instances: a novel polynomial-time approximation algorithm with provable error bound and a practical hill-climbing heuristic. We conduct detailed experiments on synthetic and real data crawled from the web to evaluate the efficiency and quality of our proposed algorithms, as well as show how product designers can benefit by leveraging collaborative tagging information. 
\end{abstract}

\category{H.4}{Information Systems Applications}{Miscellaneous}

\terms{Algorithms, Performance}

\keywords{collaborative tagging, product design, naive bayes, optimization}


\begin{bottomstuff}
The work of Mahashweta Das and Gautam Das is partially supported by NSF grants 0812601, 0915834, 1018865, a NHARP grant from the Texas Higher Education Coordinating Board, and grants from Microsoft Research and Nokia Research.

The work of Vagelis Hristidis is partially supported by NSF grants IIS-0811922, IIS-
0952347, HRD-0833093 and Google Research Award.

\end{bottomstuff}

\maketitle

\section{Introduction}
\label{intro}

\smallskip\noindent
{\bf Motivation:}
The widespread use and popularity of online collaborative content sites has created rich resources for users to consult in order to make purchasing decisions on various products such as movies, e-commerce products, restaurants, etc. Various websites today (e.g., Amazon for e-commerce products, Flickr for photos, YouTube for videos) encourage users to actively participate by assigning labels or {\em tag} to online resources with a purpose to promote their contents and allow users to share, discover and organize them. An increasing number of people are turning to online ratings, reviews and user-specified tags to choose from among competing products. Products with desirable tags (e.g., {\tt modern}, {\tt reliable}, etc.) have a higher chance of being selected by prospective customers. This creates an opportunity for product designers to design better products that are likely to attract desirable tags when published. In addition to traditional marketplaces like electronics, autos or apparel, tag desirability also extends to other diverse domains. For example, music websites such as {\tt Last.fm} use social tags to guide their listeners in browsing through artists and music. An artist creating a new musical piece can leverage the tags that users have selected, in order to select the piece's attributes (e.g. acoustic and audio features) that will increase its chances of becoming popular. Similarly, a blogger can select a topic based on the tags that other popular topics have received.

Our paper investigates this novel tag maximization problem, i.e., how to leverage collaborative tagging information to decide the attribute values of new products and to return the top-$k$ products that are likely to attract the maximum number of desirable tags when published. We provide more details as follows. 

\smallskip\noindent
{\bf Tag Maximization Problem:} Assume we are given a training data of objects (i.e., products), each having a set of well-defined features (i.e., attributes) and a set of user-submitted tags (e.g., cell phones on Amazon's website, each described by a set of attributes such as {\sf\small display size}, {\sf\small Operating System} and associated user tags such as {\tt lightweight}, {\tt easy to use}). From this training data, for each distinct tag, we assume a classifier has been constructed for predicting the tag given the attributes. Tag prediction is a recent area of research (see Section~\ref{rel} for discussion of related work), and the existence of such classifiers is a key assumption in our work. In addition to the product's explicitly specified attributes, other implicit factors also influence tagging behavior, such as the perceived utility and product quality to the user, the tagging behavior of the user's friends, etc. However, pure {\em content-based} tag prediction approaches are often quite effective $-$ e.g., in the context of laptops, attributes such as smaller dimensions and the absence of a built-in DVD drive may attract tags such as {\tt portable}. 

Given a query consisting of a subset of tags that are considered {\em desirable}, our task is to suggest a new product (i.e., a combination of attribute values) such that the expected number of desirable tags for this potential product is maximized. This can be extended to the top-$k$ version, where the task is to suggest the $k$ potential products with the highest expected number of desirable tags. In addition to the set of desirable tags, our problem can also consider a set of {\em undesirable} tags, e.g. {\tt unreliable}. The optimization goal in this case is to maximize the number of desirable tags and minimize the undesirable ones - a simple combination function is to optimize the expected number of desirable tags minus the expected number of undesirable tags. In our discussion so far, we have not explained how the set of desirable and undesirable tags are created. Although this is not the focus of this paper, we mention several ways in which this can be done. For example, domain experts could study the set of tags and mark them accordingly. Automated methods may involve leveraging the user rating or the sentiment of the user review to classify tags as desirable, undesirable or unimportant.

\smallskip\noindent
{\bf Novelty, Technical Challenges and Approaches:}
The dynamics of social tagging has been an active research area in recent years. However related literature primarily focuses on the problems of tag prediction, including cold-start recommendation to facilitate web-based activities. To our best knowledge, tags have not been studied in the context of product design before. Of course, real-world product design is a complex task, and is an area that has been heavily studied in economics, marketing, industrial engineering and more recently in computer science. Many factors like the cost and return on investment are currently considered. We argue that the user feedback (in the form of tags of existing competing products) should be taken into consideration in the design process, especially since online user tagging is extremely widespread and offers unprecedented opportunities for understanding the collective opinion and preferences of a huge consumer base. We envision user tags to be one of the several factors in product design that can be used in conjunction with more traditional factors - e.g., our algorithms return $k$ potential new products that maximize the number of desirable tags; and this information can assist content producers, who can then further post-process the returned results using additional constraints such as profitability, price, resource constraints, product diversity, etc. Moreover, product designers can explore the data in an interactive manner by picking and choosing different sets of desirable tags to get insight on how to build new products that target different user populations $-$ e.g., in the context of cell phones, tags such as {\tt lightweight} and {\tt powerful} target professionals, whereas tags such as {\tt cheap}, {\tt cool} target younger users.

Solving the tag maximization problem is technically challenging. In most product bases, complex dependencies exist among the tags and products, and it is difficult to determine a combination of attribute values that maximizes the expected number of desirable tags. In this paper we consider the very popular Naive Bayes Classifier for tag prediction \footnote{Naive Bayes Classifiers are often effective, rival the performance of more sophisticated classifiers, and are known to perform well in social network applications. For instance, Pak and Paroubek~\cite{twitter2010} show that Naive Bayes performs better than SVM and CRF in classifying the sentiment of blogs.}. Extending our work for other popular classifiers is one of our future research directions. As one of our first results, we show that even for this classifier (with its simplistic assumption of conditional independence), the tag maximization problem is NP-Complete. Given this intractability result, it is important to develop algorithms that work well in practice.  A highlight of our paper is that we have avoided resorting to heuristics, and instead have developed principled algorithms that are practical {\em and at the same time} possess compelling theoretical characteristics. We also mention a practical heuristic that works very well for real-world instances.

Our first algorithm is a novel exact top-$k$ algorithm \textbf{ETT} (Exact Two-Tier Top-$k$ algorithm) that performs significantly better than the naive brute-force algorithm (which simply builds all possible products and determines the best ones), for moderate problem instances. Our algorithm is based on nontrivial adaptations of top-$k$ query processing techniques (e.g., ~\cite{topk2001}),  but is not merely a simple extension of TA. The complexity arises because the problem involves maximizing a sum of terms, where within each term there is a product of quantities which are interdependent with the quantities from the other terms. Our top-$k$ algorithm and has an interesting two-tier architecture. At the bottom tier, we develop a sub-system for each distinct tag, such that each sub-system has the ability to compute on demand a stream of products in order of decreasing probability of attracting the corresponding tag, {\em without having to pre-compute} all possible products in advance. In effect, each sub-system simulates {\em sorted access} efficiently. This is achieved by partitioning the set of attributes into smaller groups (thus, each group represents a {\em partial} product), and running a separate merge algorithm over all the groups. The top tier considers the products retrieved from each sub-system in a round-robin manner, computes the expected number of desirable tags for each retrieved product, and stops when a threshold condition is reached. Although in the worst case this algorithm can take exponential time, for many datasets with strong correlations between attributes and tags, the stopping condition is reached much earlier.

However, although the exact algorithm performs well for moderate problem sizes, it did not easily scale to larger real-world sized datasets, and thus we also develop several approximation algorithms for solving the problem. Designing approximation algorithms with guaranteed behavior is challenging, since no known approximation algorithm for other NP-Complete problems can be easily modified for our case. Our exact algorithm ETT can be modified to serve as an approximation algorithm - we can {\em change the threshold condition} such that the algorithm stops when the threshold is within a small user-provided approximation factor of the top-$k$ product scores produced thus far. This algorithm can guarantee an approximation factor in the quality of products returned, but would run in exponential time in the worst case. Our first approximation algorithm \textbf{PA} (Poly-Time Approximation algorithm) runs in worst case polynomial time, {\em and} also guarantees a provable bound on the approximation factor in product quality. The principal idea is to group the desirable tags into {\em constant-sized} groups, find the top-$k$ products for each sub-group, and output the overall top-$k$ products from among these computed products. Interestingly, we note that in this algorithm we create sub-problems by grouping tags; in contrast in our exact algorithm we create sub-problems (i.e., subsystems) by grouping attributes. For each sub-problem thus created, we show that it can be solved by a polynomial time approximation scheme (PTAS) given any user-defined approximation factor. The algorithm's overall running time is exponential only in the (constant) size of the groups, thus giving overall a polynomial time complexity. 

Our second approximation algorithm is a more practical hill climbing {\em heuristic} \textbf{HC}. It starts with a randomly generated product (starts at the base of the hill), and then repeatedly improves the solution by changing some attribute values (walks up the hill) until some no further small changes improves the product (reaches a local maximum). This algorithm can be improved by repeating with random restarts. Multiple ($k$ or more) random restarts may also lead us to the multiple locally optimal products, of which the top-$k$ may be returned. This algorithm works well in practice, as shown empirically in Section~\ref{expt}. But we propose this as a viable efficient solution to the problem for handling large real datasets; it does not guarantee any sort of worst case behavior, either in running time or in product quality. In fact, we prove that there exist datasets for which the expected number of tags of a globally optimum product can be exponentially larger than that of a locally optimum product. 



We experiment with synthetic as well as real datasets crawled from the web to compare our algorithms. User and case study on the real dataset demonstrates that products suggested by our algorithms appear to be meaningful. With regard to efficiency, the exact algorithm performs well on moderate problem instances, whereas the approximation algorithms scaled very well for larger datasets.

\smallskip\noindent
{\bf Summary of Contributions:}
We make the following main contributions.
\begin{itemize}
\item We introduce the novel problem of top-$k$ product design based on user-submitted tags and show that this problem is NP-complete, even if tag prediction is modeled using simple Naive Bayes Classifiers.
\item We develop an exact algorithm ETT to compute the top-$k$ best products that works well for moderate problem instances.
\item We also present a set of approximation algorithms for larger problem instances: HC, empirically shown to work extremely well for real datasets; and PA based on a polynomial time approximation scheme (PTAS), with provable error bounds. 
\item We perform detailed experiments on synthetic and real datasets crawled from the web to demonstrate the effectiveness of our developed algorithms.
\end{itemize}


\section{Problem Framework}
\label{prelim}

Let $\mathbb{D}$ = \{$\mathnormal{o_1}$, $\mathnormal{o_2}$, ..., $\mathnormal{o_n}$\} be a collection of $\mathnormal{n}$ products, where each product entry is defined over the attribute set $\mathnormal{A}$ = \{$\mathnormal{A_1}$, $\mathnormal{A_2}$, ..., $\mathnormal{A_m}$\} and the tag dictionary space $\mathnormal{T}$ = \{$\mathnormal{T_1}$, $\mathnormal{T_2}$, ..., $\mathnormal{T_r}$\}. Each attribute $\mathnormal{A_i}$ can take one of several values $\mathnormal{a_i}$ from a multi-valued categorical domain $\mathnormal{D}_i$, or one of two values {0, 1} if a boolean dataset is considered. The attribute set $\mathnormal{A}$ can be a mix of categorical and boolean attributes too.
A tag $\mathnormal{T_j}$ is a bit where a 0 implies the absence of a tag and a 1 implies the presence of a tag for product $\mathnormal{o}$. Each product is thus a vector of size ($\mathnormal{m}$  + $\mathnormal{r}$), where the first $\mathnormal{m}$ positions correspond to a vector of attribute values, and the next $\mathnormal{r}$ positions correspond to a boolean vector.\footnote{Our framework allows numeric attributes, but as is common with Naive Bayes Classifiers, we assume that they have been appropriately binned into discrete ranges.}


\begin{example}
\textup{
Consider a camera dataset with $\mathnormal{n}=3$ rows, $\mathnormal{m}=4$ attributes and $\mathnormal{r}=3$ tags, where each tuple represents a camera. The categorical attributes are {\sf\small Brand}, {\sf \small Type}, etc., and the boolean attributes are {\sf\small Auto Focus}, {\sf\small Image Stabilizer}, etc. Suppose there are three user-submitted tags namely , {\tt lightweight}, {\tt user-friendly}, and {\tt excellent quality}. The value of a tag column is 1 if the camera entry tuple has been annotated by this tag, and 0 otherwise. An example of such a camera dataset, having a mix of categorical and boolean attributes is shown in Table~\ref{table:database}.  A camera manufacturing company may investigate such a training set to learn the correlation between attributes and tags and design {\em new camera(s)} with the {\em best} attribute values so that it generates maximum positive response from the customers. $\Box$ }
\end{example}

\vspace{-0.15in}
\begin{table*}[!htb]
\centering
\tbl{Sample camera training set of boolean and categorical attributes, as well as user-submitted tags}{
\vspace{0.05in}
\begin{tabular}{|c|c|c|c|c|c|c|c|c|} \hline
\multicolumn{1}{|c|}{}&\multicolumn{4}{|c|}{Attribute}&\multicolumn{3}{|c|}{Tag}\\
\hline
ID&Brand&Type&Auto Focus&Image Stabilizer&lightweight&user-friendly&excellent quality\\ \hline
1&Nikon&SLR&1&1&0&0&1\\ \hline
2&Canon&Compact&1&1&1&1&0\\ \hline
3&Sony&Compact&1&0&1&0&0\\ \hline
\end{tabular}}
\label{table:database}
\end{table*}

We assume such a dataset has been used as a training set to build Naive Bayes Classifiers (NBC), that classify tags given attribute values (one classifier per tag). The classifier for tag $\mathnormal{T_j}$ defines the probability that a new product $\mathnormal{o}$ is annotated by tag $\mathnormal{T_j}$:

\vspace{-0.15in}
\begin{eqnarray}
 Pr( \mathnormal{T_j} \mid \mathnormal{o} ) &=& Pr( \mathnormal{T_j} \mid \mathnormal{a_1}, \mathnormal{a_2}, ..., \mathnormal{a_m} )\nonumber\\
 &=& \frac{Pr( \mathnormal{T_j}) . \Pi_{i=1}^{\mathnormal{m}} Pr( \mathnormal{a_i} \mid \mathnormal{T_j} )}{Pr(\mathnormal{a_1}, \mathnormal{a_2}, ..., \mathnormal{a_m} )}
\label{eq1}
\end{eqnarray}

\noindent
where $\mathnormal{a_i}$ is the value of $\mathnormal{o}$ for attribute $\mathnormal{A_i}$, $\mathnormal{a_i}$ $\epsilon$ $\mathnormal{D}_i$.
The probabilities $Pr( \mathnormal{a_i} \mid \mathnormal{T_j} )$ are computed using the dataset. In particular, $Pr( \mathnormal{a_i} \mid \mathnormal{T_j} )$ is the proportion\footnote{The observed probabilities
are smoothened using the Bayesian $m$-estimate method~\cite{cestnik90}. We note that more sophisticated Bayesian methods that use an informative prior may be employed instead.} of products tagged by $\mathnormal{T_j}$ that have $\mathnormal{A_i}=\mathnormal{a_i}$. $Pr(\mathnormal{T_j})$ is the proportion of products in the dataset that has $\mathnormal{T_j}$.

Similarly, we compute the probability $Pr(\mathnormal{T_j}^\prime \mid \mathnormal{o})$ of a product $\mathnormal{o}$ not having tag $\mathnormal{T_j}$:

\vspace{-0.15in}
\begin{eqnarray}
Pr( \mathnormal{T_j}^\prime \mid \mathnormal{o} ) &=& \frac{Pr( \mathnormal{T_j}^\prime) . \Pi_{i=1}^{m} Pr( \mathnormal{a_i} \mid \mathnormal{T_j}^\prime )}{Pr(\mathnormal{a_1}, \mathnormal{a_2}, ..., \mathnormal{a_m} )}
\label{eq2}
 \end{eqnarray}

We know that $Pr(\mathnormal{T_j} \mid \mathnormal{o}) + Pr(\mathnormal{T_j}^\prime \mid \mathnormal{o})=1 $; hence from Equations~\ref{eq1}, \ref{eq2} we get $\colon$

\vspace{-0.15in}
\begin{eqnarray}
Pr(\mathnormal{a_1}, \mathnormal{a_2}, ..., \mathnormal{a_m} ) &=& Pr( \mathnormal{T_j}) . \Pi_{i=1}^{\mathnormal{m}} Pr( \mathnormal{a_i} \mid \mathnormal{T_j} ) +\nonumber\\
   && Pr( \mathnormal{T_j}^\prime) . \Pi_{i=1}^{\mathnormal{m}} Pr( \mathnormal{a_i}  \mid \mathnormal{T_j}^\prime )
\label{eq3}
\end{eqnarray}

From Equations~\ref{eq1} and \ref{eq3}:
\small
\begin{eqnarray}
Pr( \mathnormal{T_j} \mid \mathnormal{o} ) &=& Pr( \mathnormal{T_j} \mid \mathnormal{a_1}, \mathnormal{a_2}, ..., \mathnormal{a_m} )\nonumber\\
&=& \frac{Pr( \mathnormal{T_j}) . \Pi_{i=1}^{\mathnormal{m}} Pr( \mathnormal{a_i} \mid \mathnormal{T_j} )}{Pr( \mathnormal{T_j}) . \Pi_{i=1}^{\mathnormal{m}} Pr( \mathnormal{a_i} \mid \mathnormal{T_j} ) + Pr( \mathnormal{T_j}^\prime) . \Pi_{i=1}^{\mathnormal{m}} Pr( \mathnormal{a_i} \mid \mathnormal{T_j}^\prime )}\nonumber\\
&=& \frac{1}{1 + \frac{Pr( \mathnormal{T_j}^\prime)}{Pr( \mathnormal{T_j})}  \Pi_{i=1}^{\mathnormal{m}} \frac {Pr( \mathnormal{a_i} \mid \mathnormal{T_j}^\prime )}{Pr( \mathnormal{a_i} \mid \mathnormal{T_j} )}}\nonumber
\end{eqnarray}
\normalsize

For convenience we use the notation
\begin{equation}
R_j = \frac{Pr( \mathnormal{T_j}^\prime)}{Pr( \mathnormal{T_j})}  \Pi_{i=1}^{\mathnormal{m}} \frac {Pr( \mathnormal{a_i} \mid \mathnormal{T_j}^\prime )}{Pr( \mathnormal{a_i} \mid \mathnormal{T_j} )}
\label{eqr}
\end{equation}

Consider a query which picks a set of desirable tags $T^d=\{T_1,\dots,T_z\} \subseteq T$.  

The expected number of desirable tags $\mathnormal{T_j}\in \mathnormal{T^d}$ that a new product $\mathnormal{o}$, characterized by ($\mathnormal{a}_1$, $\mathnormal{a}_2$, ..., $\mathnormal{a}_m) \in \mathnormal{A}$ is annotated with, is given by$\colon$

\vspace{-0.15in}
\begin{eqnarray}
\mathbb{E}(\mathnormal{o}, \mathnormal{T}^d)=\Sigma_{j=1}^ \mathnormal{z} \frac{1}{1 + \mathnormal{R_j}}
\label{eqf}
\end{eqnarray}

We are now ready to formally define the main problem.

\vspace{0.10in}
{\bf TAG MAXIMIZATION PROBLEM}:
{\em Given a dataset of tagged products $\mathbb{D}$ = \{$\mathnormal{o_1}$, $\mathnormal{o_2}$, ..., $\mathnormal{o_n}$\}, and a query $T^d$, design $k$
new products that have the highest expected number of desirable tags they are likely to receive, given by Equation~\ref{eqf}.}

\smallskip
For the rest of the paper, we consider boolean attributes so that for attribute $\mathnormal{A_i}$, its value $\mathnormal{a_i}$ is either 0 or 1. We explain our algorithms in a boolean framework, which can be readily generalized to handle categorical data. We also assume that all tags are of equal {\em weight}$-$ if tags are of varying importance, Equation~\ref{eqf} can be re-written as a weighted sum, and all our proposed algorithms can be modified accordingly.

We now analyze the computational complexity of the main problem and then propose our algorithmic solutions.

\section{Complexity Analysis}
\label{comp}

In this section, we analyze the computational complexity of the main problem. Clearly, the brute-force exhaustive search will require us to design all possible $2^{\mathnormal{m}}$ number of products and compute $\mathbb{E}$($\mathnormal{o}$, $\mathnormal{T}^d$) for each of them. This naive approach will run in exponential time. However, we next give a proof sketch that the Tag Maximization problem is NP-Complete, which leads us to believe that in the worst case we may not be able to do much better than the naive approach.

\begin{theorem}
The Tag Maximization problem is NP-Complete even for boolean datasets and for $k = 1$.
\end{theorem}

\vspace{-0.05in}
\textit{Proof}$\colon$ The membership of the decision version of the problem in NP is obvious. To verify NP-hardness, we reduce the 3SAT problem to the decision version of our problem. We first reduce the 3SAT problem to the minimization version of the optimization problem, represented as $\mathbb{E}^{min}$($\mathnormal{o}$, $\mathnormal{T}^d$) and then reduce $\mathbb{E}^{min}$($\mathnormal{o}$, $\mathnormal{T}^d$) to $\mathbb{E}$($\mathnormal{A}$, $\mathnormal{T}^d$).

\vspace{0.05in}
Reduction of 3SAT to decision version of $\mathbb{E}^{min}$($\mathnormal{o}$, $\mathnormal{T}^d$)$\colon$

3SAT is the popular NP-complete boolean satisfiability problem in computational complexity theory, an instance of which concerns a boolean expression in conjunctive normal form, where each clause contains exactly 3 literals. Each clause $\mathnormal{C}_j$ is mapped to a tag $\mathnormal{T}_j$ in the instance of $\mathbb{E}^{min}$($\mathnormal{o}$, $\mathnormal{T}^d$) and each variable $\mathnormal{x}_i$ is mapped to attribute value $\mathnormal{a}_i$. We make the following assignments so that if there is a boolean assignment vector $\vec{\mathnormal{a}}$ = [$\mathnormal{a}_1$, ..., $\mathnormal{a}_m$] that satisfies 3SAT, then $\mathbb{E}^{min}$($\mathnormal{o}$, $\mathnormal{T}^d$) equals zero (and if $\vec{\mathnormal{a}}$ does not satisfy 3SAT,  then $\mathbb{E}^{min}$($\mathnormal{o}$, $\mathnormal{T}^d$) has a non-zero sum).

\begin{itemize}
\item {For a variable $\mathnormal{x}_i$ specified as positive literal in 3SAT, set Pr($\mathnormal{a_i}$ = 0 $\mid$ $\mathnormal{T_j}$) = 1}
\item {For a variable $\mathnormal{x}_i$ specified as negative literal in 3SAT, set Pr($\mathnormal{a_i}$ = 1 $\mid$ $\mathnormal{T_j}$) = 1}
\item {For a particular clause and for the unspecified attributes (variables), set Pr($\mathnormal{a_i}$ = 0 $\mid$ $\mathnormal{T_j}$) = Pr($\mathnormal{a_i}$ = 1 $\mid$ $\mathnormal{T_j}$) = 1}
\end{itemize}

For example, consider 3SAT instance $(\neg x_1 \vee x_2 \vee \neg x_3) \wedge (x_1 \vee \neg x_2 \vee \neg x_4)$. For each tag, we create two products. For the first clause (that corresponds to the first tag), $x_1$ (that corresponds to $A_1$) is negative and hence for both the first and second product it is $A_1=1$. $x_4$ is missing from the first clause; hence for the first product it is $A_4=0$ and for the second it is $A_4=1$. Similarly, the assignments of the second clause (that is the second tag) can be explained. Again, an assignment : $A_1=1, A_2=1, A_3=0, A_4=0$  satisfying the 3SAT instance has $\mathbb{E}^{min}$($\mathnormal{o}$, $\mathnormal{T}^d$) = 0. 

\begin{table}[!thb]
\centering
\tbl{Table of attributes and tags}{
\vspace{0.05in}
\begin{tabular}{|c|c|c|c|c|c|} \hline
\multicolumn{4}{|c|}{Attributes}&\multicolumn{2}{|c|}{Tags}\\
\hline
$A_1$&$A_2$&$A_3$&$A_4$&$T_1$&$T_2$\\ \hline
1&0&1&0&1&0\\ \hline
1&0&1&1&1&0\\ \hline
0&1&0&1&0&1\\ \hline
0&1&1&1&0&1\\ \hline
\end{tabular}}
\label{table:npcdatabase}
\end{table}

Reduction of $\mathbb{E}^{min}$($\mathnormal{A}$, $\mathnormal{T}^d$) to $\mathbb{E}$($\mathnormal{A}$, $\mathnormal{T}^d$) $\colon$

If we have a boolean assignment vector $\vec{\mathnormal{a}}$ = [$\mathnormal{a}_1$, ..., $\mathnormal{a}_m$] that minimizes the expected number of tags being present, we have the corresponding {Pr($\mathnormal{T_j}^\prime$ $\mid$ $\mathnormal{a_1}$, $\mathnormal{a_2}$, ..., $\mathnormal{a_m}$)}.
Hence, we get Pr($\mathnormal{T_j}$ $\mid$ $\mathnormal{a_1}$, $\mathnormal{a_2}$, ..., $\mathnormal{a_m}$) = 1 - Pr($\mathnormal{T_j}^\prime$ $\mid$ $\mathnormal{a_1}$, $\mathnormal{a_2}$, ..., $\mathnormal{a_m}$) that maximizes the expected number of tags being present. $\Box$

Section~\ref{exact} and Section~\ref{app} next describe our algorithmic solutions to this NP-Complete problem in a boolean framework.

\section{Exact Algorithms}
\label{exact}

A brute-force exhaustive approach (henceforth, referred to as {\bf Naive}) to solve the Tag Maximization problem requires us to design all possible $2^{\mathnormal{m}}$ number of products and compute $\mathbb{E}$($\mathnormal{o}$, $\mathnormal{T}^d$) for each possible product. Note that the number of products in the dataset is not important for the execution cost, since an initialization step can calculate all the conditional tag-attribute probabilities by a single scan of the dataset. The Naive approach will clearly run in exponential time, and the NP-completeness proof leads us to believe that in the worst case we may not be able to do much better. Although general purpose pruning-based optimization techniques (such as branch-and-bound algorithms) can be used to solve the problem more efficiently than Naive, such approaches are only limited to constructing the top-1 product, and it is not clear how they can be easily extended for $k > 1$.

In the following subsection, we propose a novel exact algorithm for any $k$ based on interesting and nontrivial adaptations of top-$k$ query processing techniques. This algorithm is shown in practice to explore far fewer product candidates than Naive, and works well for moderate problem instances.

\subsection{Exact Two-Tier Top-k Algorithm}
\label{optalgo1}

\begin{figure}[!htb]
\centering
\includegraphics[width=3.8in,height=2.6in]{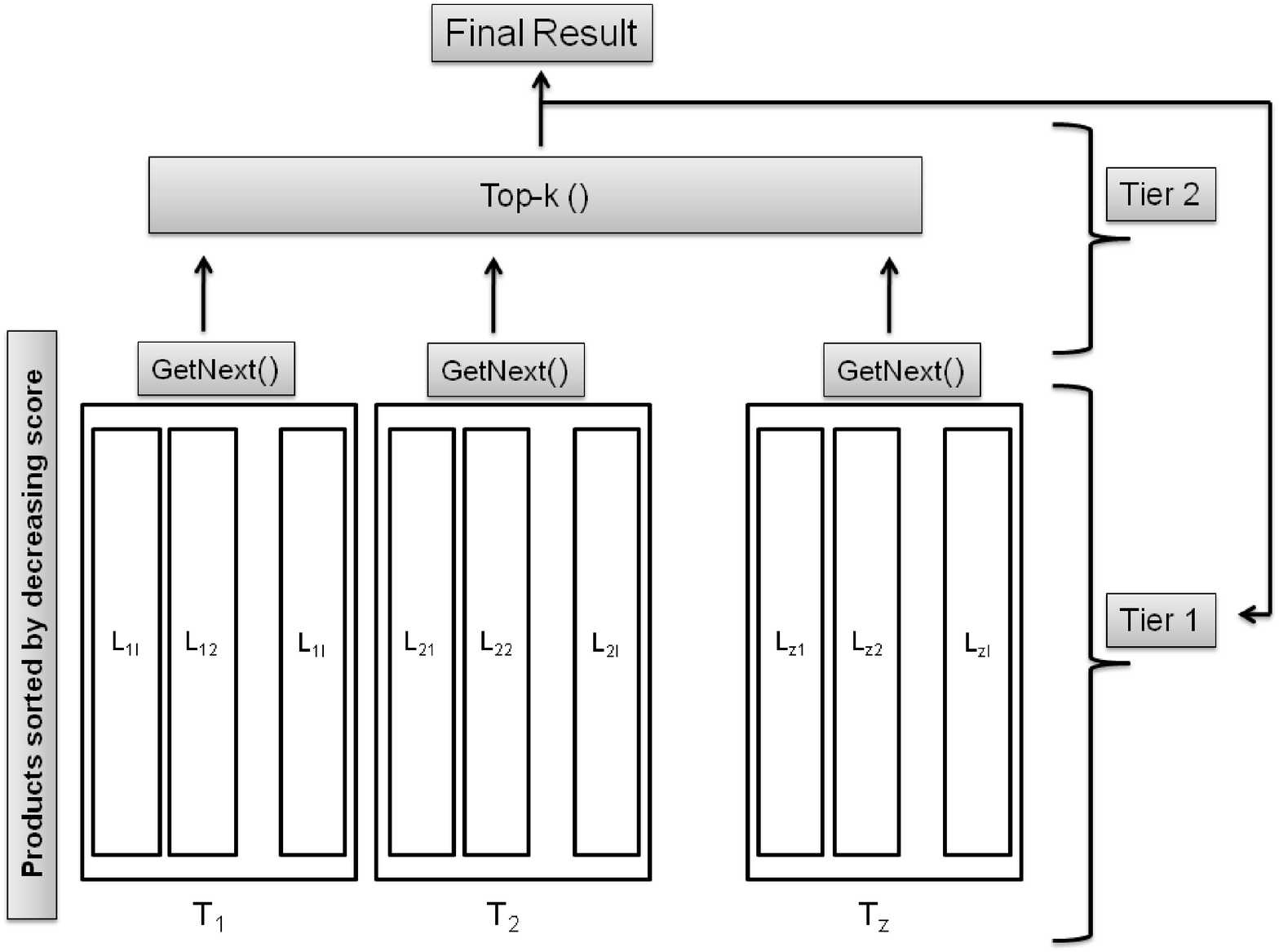}
\caption{Two-Tier Top-K Algorithm Framework}
\label{figure:ettframework}
\end{figure}

We develop an exact {\em two tier} top-$k$ algorithm ({\em ETT}) for the Tag Maximization problem.
For simplicity, henceforth we refer to desirable tags as just tags.
The main idea of {\em ETT} is to determine the {\em best} products for each individual tag in tier-1 and then match these products in tier-2 to compute the globally best products (across all tags).
Both tiers use pipelined techniques to minimize the amount of accesses, as shown in Figure~\ref{figure:ettframework}. The output of tier-1 is $z$ unbounded buffers (one for each tag) of complete products, ordered by decreasing probability for the corresponding tag. These buffers are not fully materialized, but may be considered as {\em sub-systems} that can be accessed on demand in a pipelined manner.

In tier-2, the top products from the $z$ buffers are combined in a pipelined manner to produce the global top-$k$ products, akin the {\em Threshold Algorithm} (TA)~\cite{topk2001}.
In turn, tier-2 makes GetNext() requests (see Figure~\ref{figure:ettframework}) to various buffers in tier-1 in round-robin manner.
In tier-1, for each specific tag, we partition the set of attributes into subsets, and for each subset of attributes we precompute a list of all possible partial attribute value assignments, ordered by their {\em score} for the specific tag (the score will be defined later). The partial products are then scanned and joined, leveraging results from Rank-Join algorithms~\cite{rankjoin2009} that support top-$k$ ranked join queries in relational databases, in order to feed information to tier-2. A single GetNext() for a specific tag may translate to multiple retrievals from the lists of partial products in tier-1, which are then joined into complete products and returned.

\subsubsection{Tier-1}
Suppose we partition the $m$ attributes into $l$ subsets, where each subset has $m'= \frac{m}{l}$ attributes as follows: $\{\mathnormal{a_1}, \ldots, \mathnormal{a_{\mathnormal{m}^\prime}}\}$, $\{ \mathnormal{a_{\mathnormal{m}^\prime+1}}, \ldots, \mathnormal{a_{2\mathnormal{m}^\prime}}\}$, $\ldots$, $\{\mathnormal{a_{\mathnormal{m}-\mathnormal{m}^\prime+1}}, \ldots, \mathnormal{a_{\mathnormal{m}}}\}$.
We create partial product lists $L_{j1},\dots,L_{jl}$ for each tag $T_j$. Each list $L_{ji}$ has $2^{m'}$ entries (partial products). Consider the first list $L_{j1}$. The {\em score} of a partial product $o^p\in L_{j1}$ with attribute values $a_1,\dots,a_{m'}$ for $T_j$ is

\vspace{-0.18in}
\begin{eqnarray}
\mathbb{E}_{partial}(o^p, \{\mathnormal{T}_j\}) &=& \sqrt[\mathnormal{l}]{\mathnormal{P_j}} . \Pi_{i=1}^{\mathnormal{m}^\prime} \frac{Pr( \mathnormal{a_i} \mid \mathnormal{T_j}^\prime )}{Pr( \mathnormal{a_i} \mid \mathnormal{T_j} )}
\label{eqp}
\end{eqnarray}

\noindent
where $\mathnormal{P_j}$=$\frac{Pr(\mathnormal{T_j}^\prime)}{Pr(\mathnormal{T_j})}$.
Note that the $l$-th root of $P_j$ is used in order to distribute the effect of $P_j$ from Equation~\ref{eqr} to the $l$ lists, such that when they are combined using multiplication, we get $P_j$.

Lists $L_{jl}$ are ordered by descending $\frac{1}{E_{partial}}$, since $R_j$ appears on the denominator of Equation~\ref{eqf}.
The $\mathnormal{l}$ lists are accessed in round-robin fashion and for every combination of partial products from the lists, we join them to build a complete product and resolve its exact score by Equation~\ref{eqf}.

A product is returned as a result of GetNext() to tier-2 if its score is higher than the MPFS ({\em Maximum Possible Future Score}), which is the upper bound on the score of an unseen product. To compute MPFS, we assume that the current entry from a list is joined with the top entries from all other lists$\colon$

\vspace{-0.15in}
\begin{eqnarray}
MPFS
= \frac{1}{1 + max( (\mathnormal{s_{j1}}.\mathnormal{h_{j2}}.. \cdot\mathnormal{h_{jl}}), (\mathnormal{h_{j1}}.\mathnormal{s_{j2}}.. \cdot\mathnormal{h_{jl}}),.., (\mathnormal{h_{j1}}.\mathnormal{h_{j2}} .. \cdot\mathnormal{s_{jl}}))}
\label{mpfs}
\end{eqnarray}

\noindent
where $s_{ji}$ and $h_{ji}$ are the last seen and top entries from list $L_{ji}$ respectively.

\subsubsection{Tier-2}
In this tier, the $z$ unbounded buffers, one for each tag, are combined using the summation function, as shown in Equation~\ref{eqf}. Each product from one buffer matches exactly one entry (the identical product) from each of the other buffers.
%
Products are retrieved from each buffer using GetNext() operations, and once retrieved we directly compute its score for all other tags by running each Naive Bayes, without using the process of tier-1. A product is output if its score is higher than the threshold, which is the sum of the last seen scores from all $z$ buffers. A bounded buffer with $k$ best results so far is maintained. On termination, this buffer is returned as the top-$k$ products.

\vspace{0.01in}
\noindent
The pseudocode of ETT is shown in Algorithm~\ref{alg1}.

\begin{algorithm}[!htb]
\caption{  \textbf{ETT (Naive Bayes probabilities, attributes  per group $\mathnormal{m}^\prime$, $k$)}:  top-$k$ exact products}
\label{alg1}


\begin{algorithmic}
\STATE $ $
\STATE \textit{//Main Algorithm}
\end{algorithmic}

\begin{algorithmic}[1]
\STATE Top-$k$-Buffer $\leftarrow$ $\{ \}$
\FOR{$j$ = 1 to $\mathnormal{z}$}
\STATE $B_j$ $\leftarrow$ $\{ \}$ // unbounded buffer of candidate results-products per tag
\FOR{$i$ = 1 to $\mathnormal{l}$}
\STATE $s_{ji}$, $h_{ji}$  $\leftarrow$ top entry from list $L_{ji}$
\ENDFOR
\ENDFOR
\STATE Call Threshold()
\end{algorithmic}

\begin{algorithmic}
\STATE $ $
\STATE \textit{//Method $Threshold( )$ -- Tier-2}
\end{algorithmic}

\begin{algorithmic}[1]
\WHILE{true}
\FOR{$j$ = 1 to $\mathnormal{z}$}
\STATE ($o_j$, $score_j(o_j)$ $\leftarrow$ GetNext($j$)
\STATE ExactScore($\mathnormal{o_j}$) $\leftarrow$ Compute for $\mathnormal{o_j}$ by Equation~\ref{eqf}
\ENDFOR
\STATE Update Top-$k$-Buffer with new products if necessary
\STATE MinK $\leftarrow$ lowest score in Top-$k$ buffer
\STATE $\alpha$ $\leftarrow$ $\sum_j score_j(o_j)$ // Threshold
\IF{MinK $\geq$ $\alpha$}
\RETURN top-$k$ products
\ENDIF
\ENDWHILE
\end{algorithmic}

\begin{algorithmic}
\STATE $ $
\STATE \textit{//Method GetNext$(j): (o_j, score_j(o_j))$ -- Tier-1}
\end{algorithmic}

\begin{algorithmic}[1]
\WHILE{true}
\STATE Compute MPFS by Equation~\ref{mpfs}
\STATE // $score_j(o)$ for product $o$ is defined as $1/(1+R_j)$ ($R_j$ defined by Equation 4)
\IF {$B_j$ has an product $o$ with $score_j(o) >MPFS$}
\RETURN ($o$, $score_j(o)$) AND remove it from $B_j$
\ENDIF
\STATE {Retrieve next entry $o^p$ from a list $L_{ji}$ in round robin and advance $s_{ji}$}
\STATE {Join $o^p$ with all combinations of partial products from other lists and create all products $NewProducts$}
\STATE {Add $NewProducts$ to buffer $B_j$ of candidate results-products}
\ENDWHILE

\end{algorithmic}
\label{alg1}
\end{algorithm}

\vspace{-0.15in}
\begin{table}[!htb]
\centering
\tbl{Example products data set}{
\vspace{0.05in}
\begin{tabular}{|c|c|c|c|c|c|c|} \hline
\multicolumn{5}{|c|}{Attribute}&\multicolumn{2}{|c|}{Tag}\\
\hline
ID&$\mathnormal{A_1}$&$\mathnormal{A_2}$&$\mathnormal{A_3}$&$\mathnormal{A_4}$&$\mathnormal{T_1}$&$\mathnormal{T_1}$\\ \hline
1&0&0&0&1&0&0\\ \hline
2&0&1&0&0&0&1\\ \hline
3&0&1&0&1&0&0\\ \hline
4&0&1&1&1&1&1\\ \hline
5&1&0&0&0&1&0\\ \hline
6&1&0&0&1&0&1\\ \hline
7&1&0&1&1&1&1\\ \hline
8&1&1&0&1&0&1\\ \hline
\end{tabular}}
\label{table:egdatabase}
\end{table}

\vspace{-0.15in}
\begin{figure}[!htb]
\centering
\includegraphics[width=3.75in,height=3.0in]{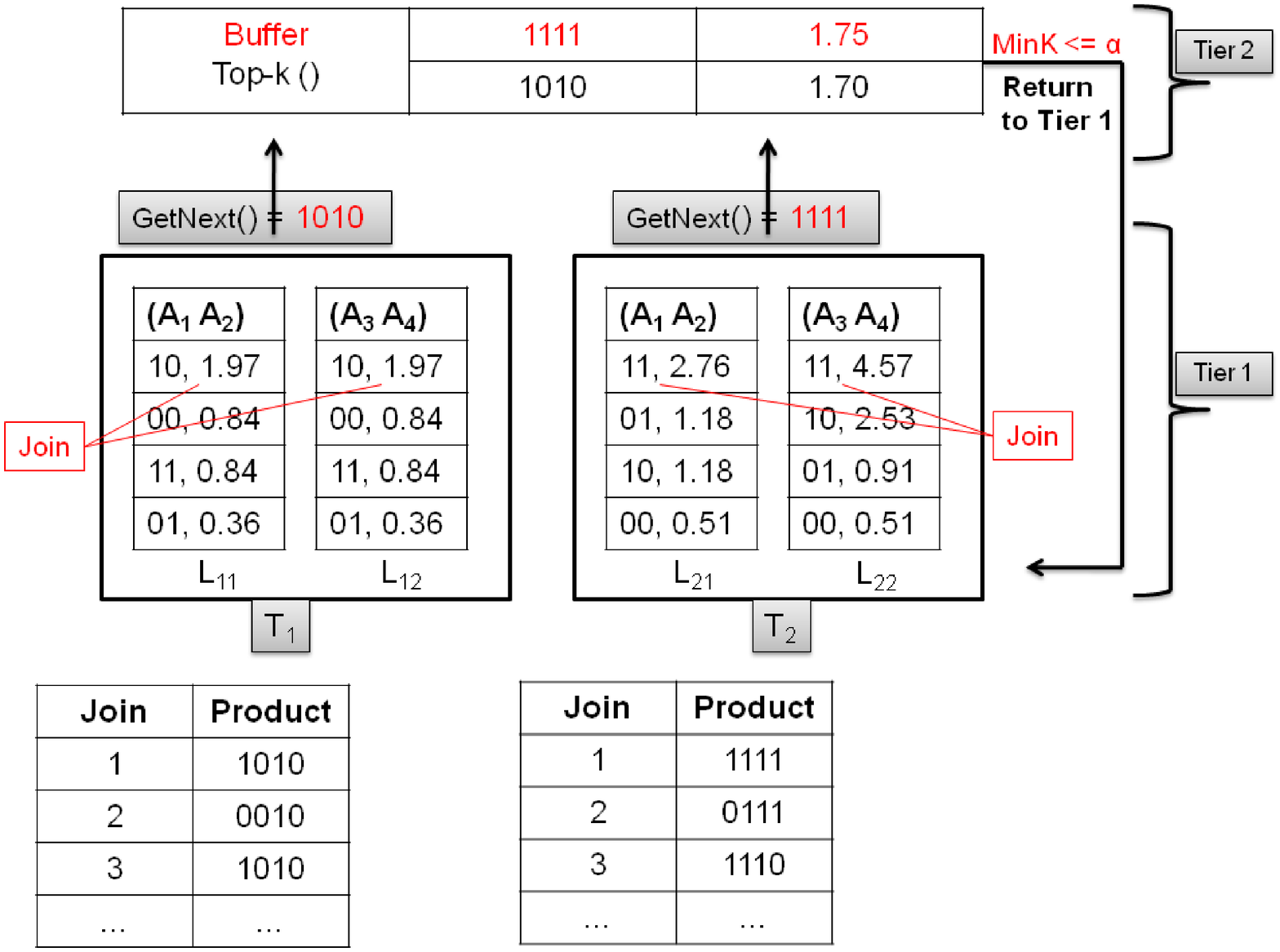}
\vspace{-0.05in}
\caption{Iteration 1: Exact Two-Tier Top-K Algorithm for Example in Table~\ref{table:egdatabase}}
\label{figure:egettrunning}
\end{figure}
\vspace{-0.10in}

\begin{example}
\label{example:egettrunning}
\textup{
Consider the boolean dataset of 10 objects, each entry having 4 attributes and 2 tags in Table~\ref{table:egdatabase}. We partition the 4 attributes into groups of 2 attributes$\colon$ ($\mathnormal{A_1}$, $\mathnormal{A_2}$) form list $\mathnormal{L_{j1}}$ and ($\mathnormal{A_3}$, $\mathnormal{A_4}$) form list $\mathnormal{L_{j2}}$. We run NBC and calculate all conditional tag-attribute probabilities. The algorithm framework for the running example is presented in Figure~\ref{figure:egettrunning}. List $\mathnormal{L_{11}}$ and $\mathnormal{L_{12}}$ under tag $\mathnormal{T_1}$ is sorted in decreasing order of $\frac{1}{E_{partial}}$, given by Equation~\ref{eqp} (similarly for $\mathnormal{L_{21}}$ and $\mathnormal{L_{22}}$ under tag $\mathnormal{T_2}$). The step-by-step operations of ETT for retrieving the top-1 product for this example is shown below $\colon$
}
\begin{enumerate}
\item{[\textbf{ITERATION  1}] Call to Threshold() in tier-2 calls GetNext() for $\mathnormal{T_1}$ and $\mathnormal{T_2}$ respectively in tier-1.}
\vspace{0.05in}
\begin{itemize}
\item{GetNext($\mathnormal{T_1}$) returns (1010,0.95) to tier-2$\colon$ Join-1 builds product $1010$, whose $score_1$(1010)=0.95 and MPFS(1010)=0.95. Since $score_{1}\geq$ MPFS, 1010 is returned.}
\item{GetNext($\mathnormal{T_2}$) returns (1111,0.93) to tier-2.}
\item{Threshold(): ExactScore(1010)=1.70, ExactScore(1111)=1.75.}
\item{Bounded Buffer$\colon$1111; MinK=1.75, $\alpha$=1.88}
\item{MinK $\leq$ $\alpha$, continue.}
\end{itemize}
\vspace{0.05in}
\item{[\textbf{ITERATION  2}] Threshold() in tier-2 calls GetNext() for $\mathnormal{T_1}$ and $\mathnormal{T_2}$ respectively in tier-1.}
\vspace{0.05in}
\begin{itemize}
\item{GetNext($\mathnormal{T_1}$) returns (1011,0.92) to tier-2.}
\item{GetNext($\mathnormal{T_2}$) returns (1110,0.88) to tier-2.}
\item{Threshold(): ExactScore(1011)=1.76, ExactScore(1110)=1.77.}
\item{Bounded Buffer$\colon$1110; MinK=1.77, $\alpha$=1.79}
\item{MinK $\leq$ $\alpha$, continue.}
\end{itemize}
\vspace{0.05in}
\item{[\textbf{ITERATION  3}] Threshold() in tier-2 calls GetNext() for $\mathnormal{T_1}$ and $\mathnormal{T_2}$ respectively in tier-1.}
\vspace{0.05in}
\begin{itemize}
\item{GetNext($\mathnormal{T_1}$) returns (0010,0.89) to tier-2.}
\item{GetNext($\mathnormal{T_2}$) returns (0111,0.84) to tier-2.}
\item{Threshold(): ExactScore(0010)=1.76, ExactScore(0111)=1.77.}
\item{Bounded Buffer$\colon$1110; MinK=1.77, $\alpha$=1.74}
\item{MinK $\geq$ $\alpha$, return 1110 and terminate.}
\end{itemize}
\end{enumerate}
Thus, ETT returns the best product by just looking up 6 products, instead of 16 products (as in Naive algorithm). $\Box$
\end{example}


\smallskip\noindent
{\bf Grouping of Attributes:} The ETT algorithm partitions the set of attributes into smaller groups (each group representing a partial product), which we join to retrieve the best product to feed to tier-2. We can employ state-of-art techniques to create a graph, where each node corresponds to an attribute and an edge between two attributes is weighed by the absolute value of the correlation between them, and then perform graph clustering techniques for partitioning the attributes into as many groups as the desired number of lists. If the sets of attributes are highly correlated, such grouping of attributes would make our ETT algorithm reach the stopping condition earlier than it would if the attributes are grouped arbitrarily. 

\section{Approximation Algorithm}
\label{app}

The exact algorithm of Section~\ref{optalgo1} is feasible only for moderate instances of the Tag Maximization problem.  For larger problem instances, it is necessary to use approximation algorithms and/or heuristics to solve the problem. In this section we discuss two such algorithms: (a) an approximation algorithm that provides guarantee in the quality of the top-k results as well as running time (guaranteed polynomial time); and (b) an efficient heuristic that provides no guarantee in either quality of the best products returned or in the running time; however, this algorithm is largely of practical interest and is empirically shown to perform well in practice.

\subsection{Poly-Time Approximation Algorithm}
\label{ptas}

Our first algorithm ({\em PA}, or {\em polynomial time approximation algorithm)} is an approximation algorithm with provable error and time bound.
The main idea is to group the desirable tags into {\em constant-sized} groups of $z'$ tags each, find the top-$k$ products for each subgroup, and output the overall top-$k$ products from among these computed products\footnote{Interestingly, we note that in this algorithm we create ($z/z'$) sub-problems by grouping tags; in contrast in our exact ETT algorithm we create sub-problems (i.e., subsystems) by grouping attributes.}. For each sub-problem thus created, we show that it can be solved by a {\em polynomial time approximation scheme} (PTAS)~\cite{Garey90}, i.e., can be solved in polynomial time given any user-defined approximation factor $\epsilon$ (function of compression factor $\sigma$ and $m$; details later in Theorem~\ref{theorem:ptas}). The overall running time of the algorithm is exponential only in the (constant) size of the groups, thus giving a overall polynomial time complexity.

\begin{algorithm}
\caption{  \textbf{PA (Naive Bayes probabilities, attributes  per group $\mathnormal{z}^\prime$, compression factor $\sigma$)}: top-$1$ approximate product in polynomial time}
\label{alg3}


\begin{algorithmic}
\STATE $ $
\STATE \textit{//Main Algorithm}
\end{algorithmic}

\begin{algorithmic}[1]
\STATE Partition tags $T$ into $z/z'$ groups $T_1, \ldots, T_{z/z'}$
\FOR{$r$ = 1 to $\frac{\mathnormal{z}}{\mathnormal{z^\prime}}$}
\STATE $o_r$ $\leftarrow$ PTAS($T_r$)
\STATE Compute ExactScore($o_r$) by Equation~\ref{eqf}
\ENDFOR
\RETURN $o_r$ with max ExactScore
\end{algorithmic}

\begin{algorithmic}
\STATE $ $
\STATE \textit{//Method $PTAS(T_r): o$}
\end{algorithmic}

\begin{algorithmic}[1]
\STATE $S'_0$ $\leftarrow \{0^m\}$  // boolean vector of size $m$ with all 0's
\FOR{$i = 1$ to $m$}
\STATE $S_i$ = $S'_{i-1} \cup S''_{i-1}$ // $S''_{i-1}:$ $S'_{i-1}$ with $i$th attribute value set to 1
\STATE // Compress $S_i$ to $S'_i$ using compression factor $\sigma$
\STATE $S'_i \leftarrow \{ \}$
\REPEAT
\STATE $o$ $\leftarrow$ representative product in $S'_{i-1}$
\STATE $S'_i \leftarrow S'_i \cup \{o\}$
\STATE Delete from $S_i$ all products $o'$ such that $\forall T_j \in T_r$,\\
$|\mathbb{E}(o, \{T_j\}) - \mathbb{E}(o', \{T_j\})| \le \sigma \mathbb{E}(o, \{T_j\})$
\UNTIL{$S_i$ is empty}
\ENDFOR
\RETURN product $o$ in $S'_m$ with largest $|\mathbb{E}(o, T_r)|$
\end{algorithmic}
\end{algorithm}
\vspace{-0.05in}

We now consider a sub-problem consisting of only a constant number of tags, $z'$. We also restrict our discussion to the case $k=1$ (more general values of $k$ are discussed later). We shall design a polynomial time approximation scheme (PTAS) for this sub-problem. A PTAS is defined as follows. Let $\epsilon > 0$ be any user-defined parameter.
Given any instance of the sub-problem, let PTAS return the product $o_a$. Let the optimal product be $o_g$. The PTAS should run in polynomial time,  and $ExactScore(o_a) \ge (1-\epsilon) ExactScore(o_g)$.

In describing the PTAS, we first discuss a simple exponential time exact top-1 algorithm for the subproblem, and then show how it can be modified to the PTAS.
Given $m$ boolean attributes and $z'$ tags, the exponential time algorithm makes $m$ iterations as follows: As an initial step, it produces the set $S^u_0$ consisting of the single product $\{0^m\}$ along with its $z'$ scores, one for each tag. In the first iteration, it produces the set containing two products $S^u_1 = \{0^m, 10^{m-1}\}$ each accompanied by its $z'$ scores, one for each tag. More generally, in the $i$th iteration, it produces the set of products $S^u_i = \{ \{0, 1\}^i \times 0^{m-1}\}$ along with their $z'$ scores, one for each tag.
Each set can be derived from the set computed in the previous iteration. Once $m$ iterations have been completed, the final set $S^u_m$ contains all $2^m$ products along with their exact scores, from which the top-1 product can be returned, which is that product for which the sum of the $z'$ scores is highest.
However, this algorithm takes exponential time, as in each iteration the sets double in size.

The main idea of the PTAS is to not allow the sets to become exponential in size. This is done by {\em compressing} each set $S_i$, having the same form as $S^u_i$ and $S_i \subseteq S^u_i$, produced in each iteration to another smaller set $S'_i$, so that they remain polynomial in size. Each product entry in $S_i$
can be viewed as points in a $z'$-dimensional space,whose $z'$ co-ordinates correspond to the product scores for $z'$ individual tags respectively, by Equation~\ref{eqf}. Essentially, we use a clustering algorithm in $z'$-dimensional space. For each cluster, we pick a {\em representative product} that stands for all other products in the cluster, which are thereby deleted. The clustering has to be done in a careful way so as to guarantee that for the products that are deleted, the representative product's exact score is be close to the deleted product's exact score. Thus when the top-1 product of the final compressed set $S'_m$ is returned, its exact score should not be too different from exact score of the top-1 product assuming no compression was done.

\vspace{0.01in}
\noindent
The pseudocode of PA is shown in Algorithm~\ref{alg3}.

\begin{figure*}[!bht]
\begin{minipage}[b]{1.0\linewidth}
\centering
\includegraphics[scale=0.25]{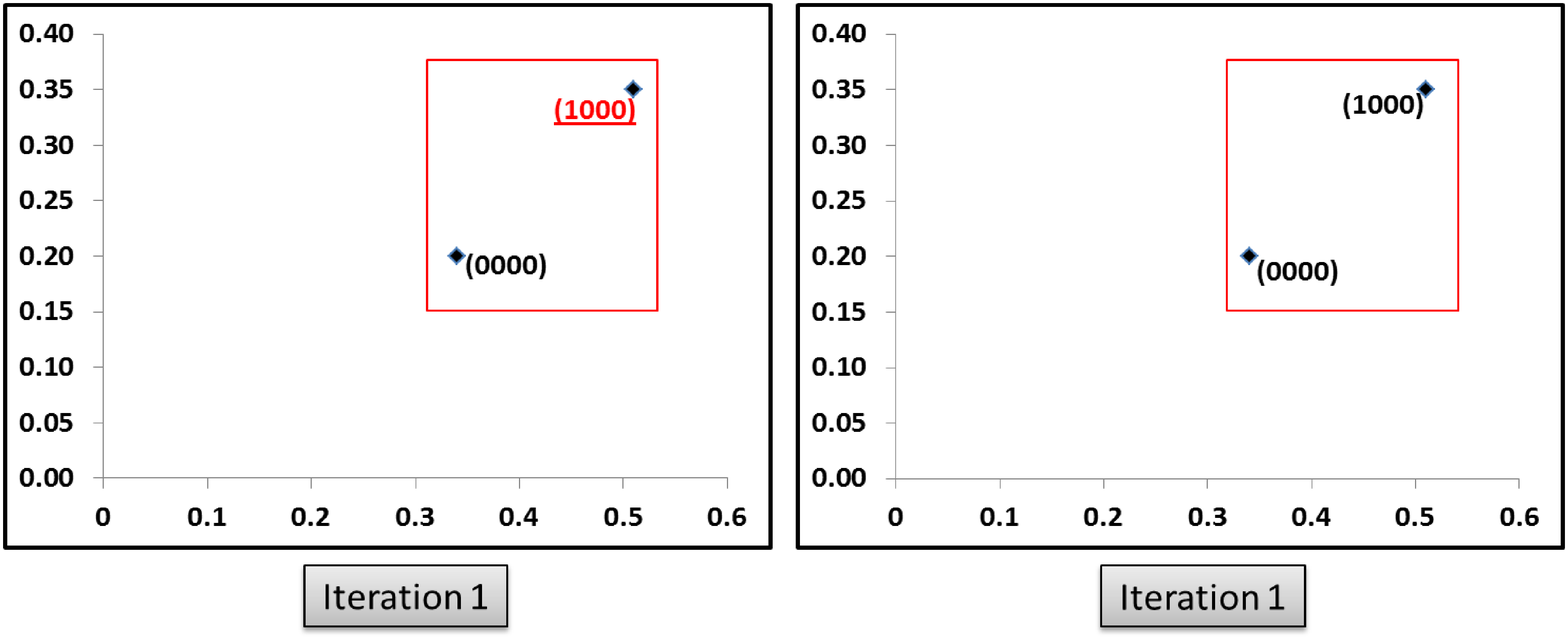}
\end{minipage}
\hspace{0.5cm}
\begin{minipage}[b]{1.0\linewidth}
\centering
\includegraphics[scale=0.25]{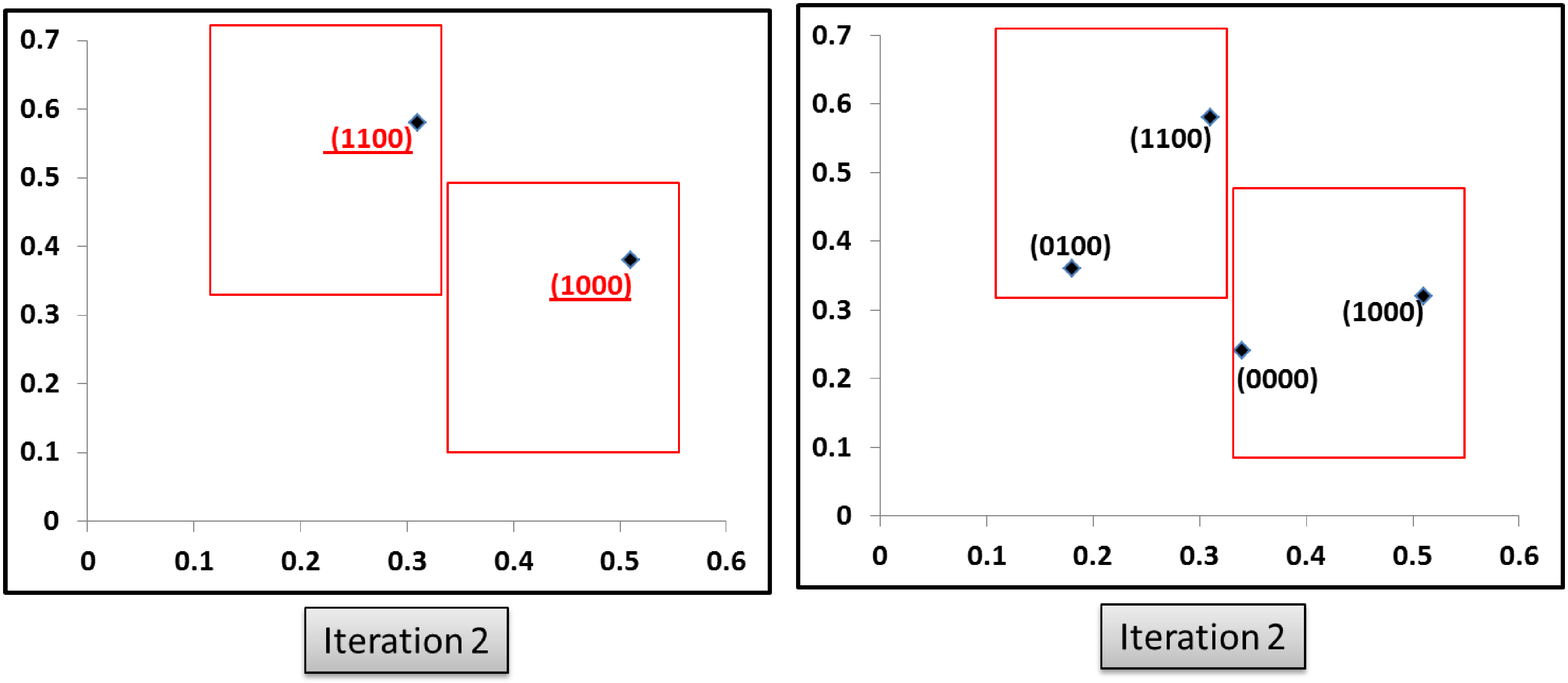}
\end{minipage}
\hspace{0.5cm}
\begin{minipage}[b]{1.0\linewidth}
\centering
\includegraphics[scale=0.25]{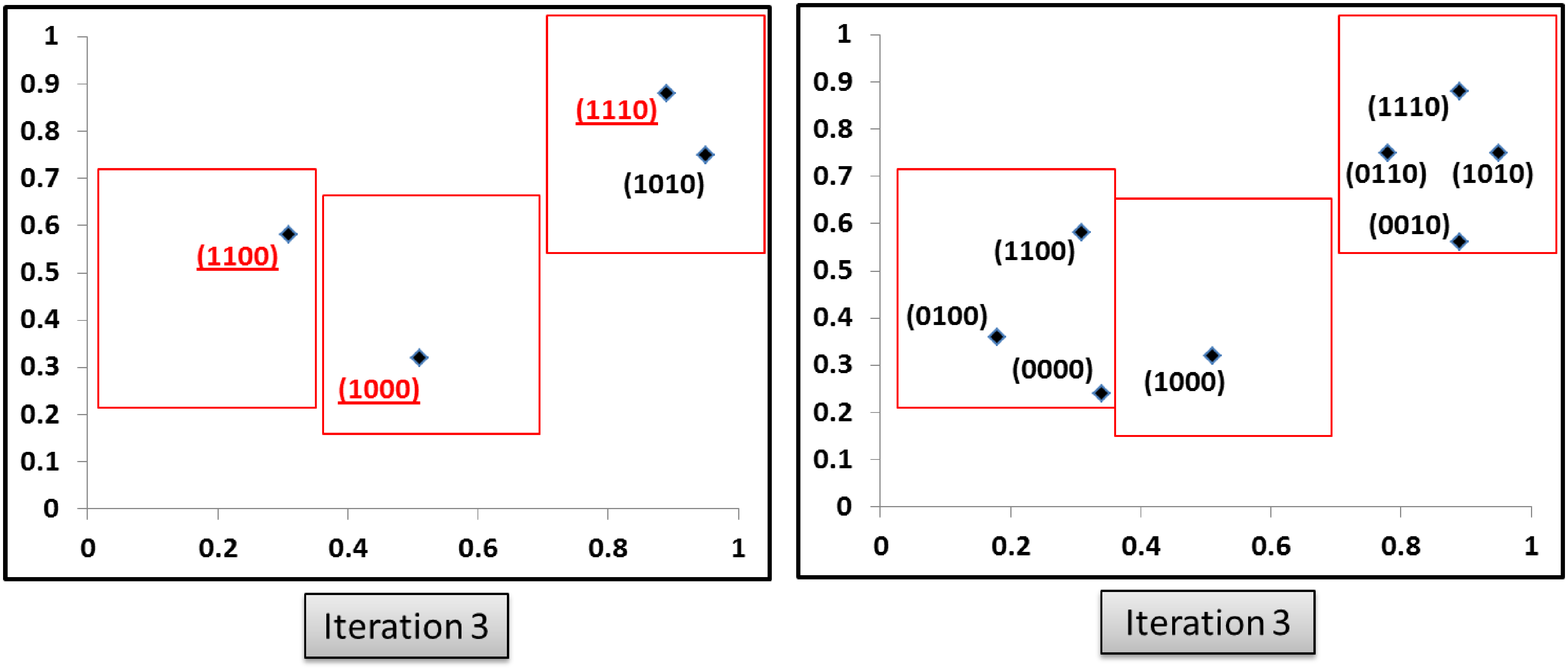}
\end{minipage}
\hspace{0.5cm}
\begin{minipage}[b]{1.0\linewidth}
\centering
\includegraphics[scale=0.25]{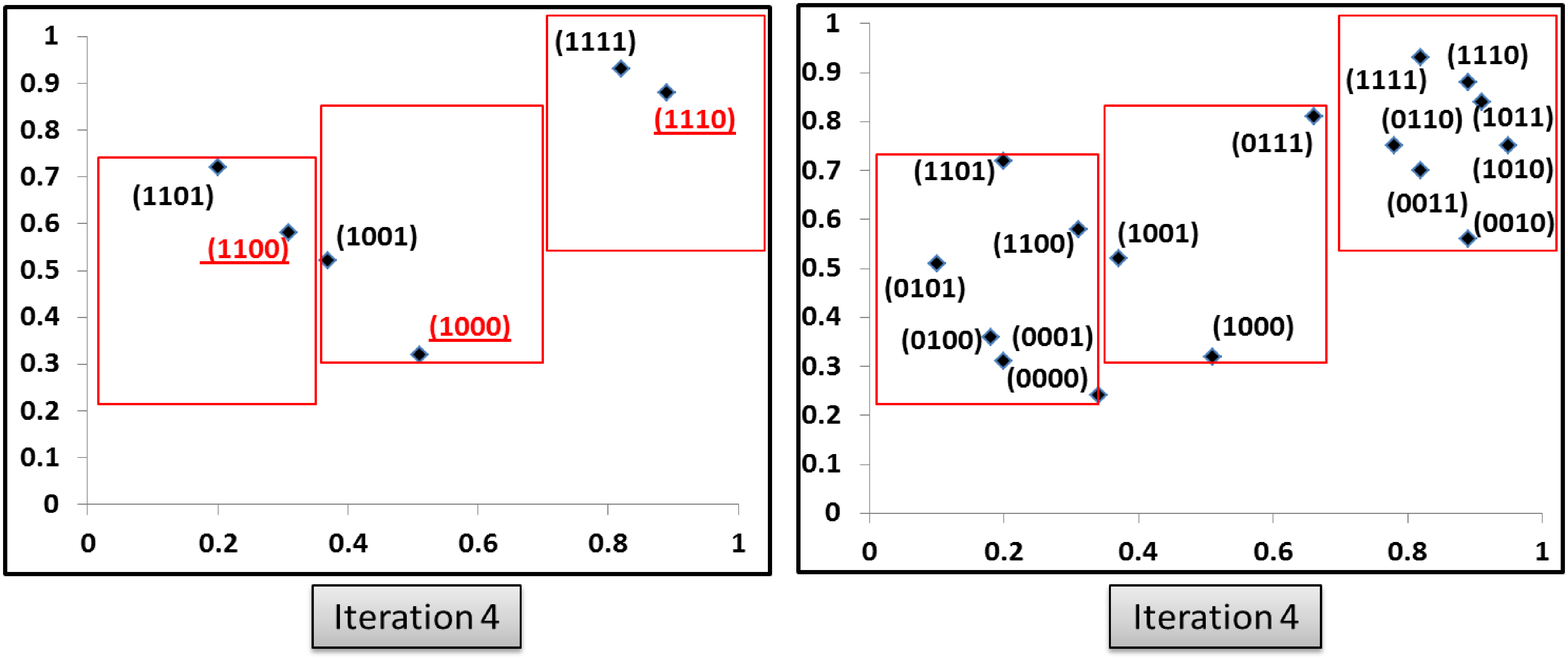}
\end{minipage}
\hspace{0.5cm}
\vspace{0.05in}
\caption{\label{figure:egptascompression}\small Compression in PA Algorithm (left column) vs. Exponential time Algorithm (right column) for Example Dataset of Two Tags in Table~\ref{table:egdatabase}}
\end{figure*}

\begin{example}
\textup{
\label{example:pa}
We execute PA on the example in Table~\ref{table:egdatabase} without any grouping of tags (i.e., $z = z' = 2$, $\frac{z}{z'}=1$), so that execution of PA is equivalent to the execution of PTAS. We also execute the exponential time top-1 algorithm (henceforth referred to as \textbf{Exponential}) which was adapted to the PTAS. Let the compression factor $\sigma$ be 0.5. We start with $S'_0 =\{0000\}$. The step-by-step operations of PA (as well as Exponential) for retrieving the top-1 product is shown below $\colon$
}
\begin{enumerate}
\item{[\textbf{ITERATION  1}]}
\vspace{0.05in}
\begin{itemize}
\item{$S_1 = \{0000, 1000\}$, each product having two-dimensional co-ordinates (0.31, 0.20) and (0.51, 0.38) respectively. For Exponential, $S^u_1 = \{0000, 1000\}$ too.}
\item{Compress $S_1$ and get $S'_1 = \{1000\}$. $1000$ is the representative product of $0000$}
\end{itemize}
\vspace{0.05in}
\item{[\textbf{ITERATION  2}]}
\vspace{0.05in}
\begin{itemize} 
\item{$S_2 = \{1000, 1100\}$ with two-dimensional co-ordinates (0.51, 0.38) and (0.31, 0.58) respectively. For Exponential, $S^u_2 = \{0000, 1000, 0100, 1100\}$ from $S^u_1$, i.e., $2^2=4$ products with two-dimensional co-ordinates (0.31, 0.20), (0.51, 0.38), (0.16, 0.38) and (0.31, 0.58) respectively.}
\item{Compress $S_2$ and get $S'_2 = \{1000, 1100\}$. In other words, no compression is possible for the $\sigma$ under consideration. }
\end{itemize}
\vspace{0.05in}
\item{[\textbf{ITERATION  3}]}
\vspace{0.05in}
\begin{itemize} 
\item{$S_3 = \{1000, 1100, 1010, 1110\}$ with two-dimensional co-ordinates (0.51, 0.38), (0.31, 0.58), (0.95, 0.75) and (0.89, 0.88) respectively from $S'_2$. For Exponential, $S^u_3 = \{0000, 1000, 0100, 1100, 0010, 1010, 0110, 1110 \}$ from $S^u_2$, i.e., $2^3=8$ products.}
\item{Compress $S_3$ and get $S'_3 = \{1000, 1100, 1110\}$. 1010 is the representative product of 1110.}
\end{itemize}
\vspace{0.05in}
\item{[\textbf{ITERATION  4}]}
\vspace{0.05in}
\begin{itemize} 
\item{$S_4 = \{1000, 1100, 1110, 1001, 1101, 1111\}$ with two-dimensional co-ordinates (0.51, 0.38), (0.31, 0.58), (0.89, 0.88), (0.37, 0.52), (0.20, 0.72) and (0.82, 0.93). For Exponential, $S^u_4$ has all $16$ products as we see in Figure~\ref{figure:egptascompression}}
\item{Compress $S_4$ and get $S'_4 = \{1000, 1100, 1111\}$.}
\end{itemize}
\end{enumerate}
The top-$1$ approximate product is $1111$ with score 1.75 while the optimal product is $1110$ with score 1.77. Figure~\ref{figure:egptascompression} shows the compression in the four iterations. The boolean products in underlined red font are the cluster representatives. $\Box$
\end{example}

\vspace{-0.05in}
\begin{theorem}
\label{theorem:ptas}
Given a user defined approximation factor $\epsilon$, a constant sized group $T_r$ of $z'$ tags, and for $k=1$, if we set the compression factor $\sigma = \epsilon/2m$, then: 
\begin{enumerate}
\item For every product $o$ in the uncompressed set $S^u_m$, there is a product $o'$ in the compressed set $S'_m$ for which $\mathbb{E}(o, T_r)\leq (1+\sigma)^m  \mathbb{E}(o', T_r)$ 
\item The output of PTAS($T_r$) has an exact score that is at least $\frac{1}{(1+\epsilon)}$ times the exact score of the optimal product
\end{enumerate}
\end{theorem}

\vspace{0.05in}
{\em Proof of Part 1}: 
Let $o^u_{i}$ indicate a product belonging to uncompressed set $S^u_i = \{\{0, 1\}^i \times 0^{m-i}\}$ in the $i^{th}$ iteration where $S^u_i$ has all $2^i$ products. Let $o_{i}$ indicate a product belonging to set $S_i$ having the same form as $S^u_i$, $S_i \subseteq S^u_i$. 
Let $o'_{i}$ indicate an product in compressed set $S'_i$, $S'_i \subseteq S_i$. Note that in the $i^{th}$ iteration, $S_i$ is built from products in the compressed set in $(i-1)^{th}$ iteration $S_{i-1}$, while $S'_i$ is built by compressing $S_i$. Intuitively, the idea is : for a single tag $T_j$, if scores of two products $o^u_{i}$ and $o'_{i}$ in $S_i$ are {\em close} to each other (so that $o^u_{i}$ is represented by $o'_{i}$ in $S'_i$, and $o^u_i$ does not exist in $S'_i$), scores of products $o^u_{i+1}$ and $o'_{i+1}$ are also close to each other, where $o^u_{i+1}$ is $o^u_i$ and $o'_{i+1}$ is $o'_i$ with $(i+1)^{th}$ bit flipped. In Example~\ref{example:pa}, product $0000$ in uncompressed set $S_1$ is represented by product $1000$ in $S'_1$ since they are close to each other; therefore, a product $0100$ in uncompressed set $S^u_2$ (but not in $S_2$ due to its removal from $S'_1$) must be close to some product belonging to $S'_2$ (which happens to be $1100$ in our example). 

More formally, we need to show that for a tag $T_j$, if $\Delta_1 = \frac{\mathbb{E}(o'_{i}, T_j)}{\mathbb{E}(o^u_{i}, T_j)} \leq (1+ \epsilon_1)$, then $\Delta_2 = \frac{\mathbb{E}(o'_{i+1}, T_j)}{\mathbb{E}(o^u_{i+1}, T_j)}  \leq (1+ \epsilon_2) \leq P(n)(1+ \epsilon_1)$, where P(n) is a polynomial in n. 

The score of product $o^u_i$ in uncompressed set $S^u_i$ for tag $T_j$ from Equation~\ref{eqf} is: 
\vspace{-0.10in}
\begin{eqnarray}
\mathbb{E}(o^u_{i}, T_j) &=& \frac{1}{1 + \frac{Pr( \mathnormal{T_j}^\prime)}{Pr( \mathnormal{T_j})}  \Pi_{i=1}^{\mathnormal{m}} \frac {Pr( \mathnormal{a_i} \mid \mathnormal{T_j}^\prime )}{Pr( \mathnormal{a_i} \mid \mathnormal{T_j})}}\nonumber\\
&=& \frac{1}{1 +  PQ} (say) \nonumber 
\end{eqnarray}

\smallskip
\noindent
where $P$ = $\frac{Pr( \mathnormal{T_j}^\prime)}{Pr( \mathnormal{T_j})}  \Pi_{1}^{\mathnormal{i}} \frac {Pr( \mathnormal{a_i} \mid \mathnormal{T_j}^\prime )}{Pr( \mathnormal{a_i} \mid \mathnormal{T_j})}$ and $Q$ =  $\frac{Pr( \mathnormal{T_j}^\prime)}{Pr( \mathnormal{T_j})}  \Pi_{i+1}^{\mathnormal{m}} \frac {Pr( \mathnormal{a_i} \mid \mathnormal{T_j}^\prime )}{Pr( \mathnormal{a_i} \mid \mathnormal{T_j})}$ are proportional to the product of probabilities for the first $i$ in $o^u_i$ and the remaining $(m-i)$ attributes in $o^u_i$ respectively. Similarly, the scores of products $o'_{i}$, $o^u_{i+1}$ and $o'_{i+1}$ for tag $T_j$ are:
\vspace{-0.10in}
\begin{eqnarray}
\mathbb{E}(o'_{i}, T_j) &=& \frac{1}{1 +  P'Q}\nonumber\\ 
\mathbb{E}(o^u_{i+1}, T_j) &=& \frac{1}{1 +  PQ'}\nonumber\\ 
\mathbb{E}(o'_{i+1}, T_j) &=& \frac{1}{1 +  P'Q'}\nonumber
\end{eqnarray}

\smallskip
\noindent
where $P'$ and $Q'$ are proportional to the product of probabilities for the first $i$ in $o'_{i}$ and the remaining $(m-i)$ attributes in $o^u_{i+1}$, $o'_{i+1}$ in which the $(i+1)^{th}$ attribute value is flipped from $o^u_{i}$. 

Assume $\mathbb{E}(o^u_i, T_j)$ is close to $\mathbb{E}(o'_{i}, T_j)$, so that $o'_{i}$ represents $o^u_{i}$ and that $P' \leq P$ so that the product of probabilities decrease (i.e., score of product increases) when the $i^{th}$ attribute value is flipped. The difference in exact score between $o^u_i$ and $o'_i$ can be expressed as: 

\vspace{-0.10in}
\begin{eqnarray}
\Delta_1 &=& \frac{\mathbb{E}(o_{i}, T_j)}{\mathbb{E}(o^u_i, T_j)}\nonumber\\
&=& \frac{1+PQ}{1+P'Q}\nonumber\\
&=& \frac{1+a}{1+b} \leq (1 + \epsilon_1) (say)
\label{eqalpha}
\end{eqnarray}

The relationship between $\mathbb{E}(o^u_{i+1}, T_j)$ and $\mathbb{E}(o'_{i+1}, T_j)$ can be similarly expressed as:

\vspace{-0.10in}
\begin{eqnarray}
\Delta_2 &=& \frac{\mathbb{E}(o'_{i+1}, T_j)}{\mathbb{E}(o^u_{i+1}, T_j)}\nonumber\\
&=& \frac{1+PQ'}{1+P'Q'}\nonumber\\
&=& \frac{ 1 + PQ {\frac{Q'}{Q}} }{ 1 + P'Q {\frac{Q'}{Q}} }\nonumber\\
&=& \frac{1+ac}{1+bc} \leq (1 + \epsilon_2) (say), \ \textnormal{where}\ c=\frac{Q'}{Q}
\label{eqbeta}
\end{eqnarray}

If 
$\epsilon_2 \leq P(n)\epsilon_1$ where $P(n)$ is a polynomial of $n$, then the proof is complete.

From Equation~\ref{eqalpha}, we get:
\vspace{-0.15in}
\begin{eqnarray}
1+a \leq (1 + \epsilon_1)(1+b) \nonumber\\
or, \frac{1+ac}{1+bc} \leq 1 + \frac{c(1+b)}{1+bc}\epsilon_1 \nonumber   
\end{eqnarray}

\smallskip
Now, $a\  \epsilon\  [\frac{1}{n^m},\ n^m ]$, $b \epsilon [\frac{1}{n^m}, n^m ]$, $c\ \epsilon\ [\frac{1}{n^2},\ n^2 ]$  so that: 
 
\vspace{-0.15in}
\begin{eqnarray}
\frac{1+b}{1+bc} \leq \frac{1+n^m}{1+n^{m+2}} \nonumber 
\end{eqnarray}

Therefore, $\epsilon_2 \leq P(n)\epsilon_1$ and $\deg P(n) = 2$. The above proof for a tag $T_j$ can be readily extended for a set of tags $T_r$. Hence in our algorithm PTAS($T_r$), each skipped product has at least one representative product retained in the compressed set.        

\vspace{0.05in}
{\em Proof of Part 2}: 
Consider any tag group $T_r$, and let $o^{OPT}$ be the optimal product for this group, and $o^{APP}$ be the product returned by PTAS. From Part 1, for every product $o$ in the set $S^u_m$ (assuming no compression was used in any iterations), there is a product $o'$ in the compressed set $S'_m$ that satisfies
\vspace{-0.08in}
\begin{eqnarray}
\mathbb{E}(o, T_r)\leq (1+\sigma)^m  \mathbb{E}(o', T_r)
\end{eqnarray}

In particular, the following holds
\vspace{-0.02in}
\begin{eqnarray}
\mathbb{E}(o^{OPT}, T_r) \leq (1+\sigma)^m \mathbb{E}(o^{APP}, T_r)
\end{eqnarray}

Since $\sigma = \epsilon/2m$, we get:

\vspace{-0.12in}
\begin{eqnarray}
\mathbb{E}(o^{OPT}, T_r)  &\leq& (1+\frac{\epsilon}{2m})^m\mathbb{E}(o^{APP}, T_r)\nonumber \\
                          &\leq& e^{\frac{\epsilon}{2}}\mathbb{E}(o^{APP}, T_r)\nonumber \\
								  &\leq& (1+\epsilon) \mathbb{E}(o^{APP}, T_r)
\end{eqnarray}

Therefore, the output of PTAS($T_r$) $o^{APP}$ has an exact score that is at least $\frac{1}{(1+\epsilon)}$ times the exact score of the optimal product $o^{OPT}$.






\begin{theorem}
\label{theorem:ptasall}
Given a user defined approximation factor $\epsilon$, a non-constant number of tags $z$ grouped into $\frac{z}{z'}$ groups of $z'$ tags per group, and for $k=1$, if we set the compression factor $\sigma = \epsilon/2m$, then: 
\begin{enumerate}
\item The output of PA has an exact score that is at least $\frac{z'}{z(1+\epsilon)}$ times the exact score of the optimal product
\item PA runs in polynomial time
\end{enumerate}
\end{theorem}

\vspace{0.05in}
{\em Proof of Part 1}: The analysis in Theorem~\ref{theorem:ptas} is for a single tag group $T_r$ having constant number of tags $z'$. Since there are $z/z'$ tag groups in totality, it is easy to see that this introduces an additional factor of $z'/z$ to the overall approximation factor, i.e., the output of PA has an exact score that is at least $\frac{z'}{z(1+\epsilon)}$ times the exact score of the optimal product.  

\vspace{0.05in}
{\em Proof of Part 2}: To show that PA is a polynomial time algorithm, the main task is to show that the compressed lists are always polynomial in length.
We first observe that probability quantities such as $Pr( \mathnormal{a_i} \mid \mathnormal{T_j} )$ are rational numbers, where both the numerator as well as the denominator are integers bounded by $n$ (i.e., the number of products in the dataset). From Equation~\ref{eqf}, note that the score of a product involves $m$ such probability quantity multiplications, where $m$ is the number of attributes. Therefore, the score of any product for any single tag can be represented as a rational number, where the numerator and denominator are integers bounded by $O(n^m)$. Thus, we can normalize each such score into an integer by multiplying it with $O(n^m)$.

Next, consider a $z'$-dimensional cube with each side of length $L = O(n^m)$.
We partition the cube into $z'$-dimensional {\em cells} as follows: Along each axis, start with the furthest value $L$, and then proceed towards the origin by marking the points $L/(1+\sigma)$, $L/(1+\sigma)^2$, and so on. The number of points marked along each axis is $\log_{(1+\sigma)} L$ = $O(m \log_{(1+\sigma)} n)$ which is a polynomial in $m$ and $n$. Then at each marked point we pass $(z'-1)$-dimensional hyperplanes perpendicular to the corresponding axis. Their intersections creates $O(poly (m, n)^{z'})$ cells within cube $L^{z'}$.

Due to this skewed method of partitioning cube into cells, we see that the cells that are further away from the origin are larger.
Consider the $i$th iteration of the PTAS algorithm. Each product in $S_i$ may be represented as a point in this cube. Though within any cell there may be several points corresponding to products of $S_i$, after compression {\em there can be at most only one} point corresponding to a product of $S'_r$, because two or more points could not have survived the compression process. 
The length of any compressed list in the PTAS algorithm is at most $O(poly (m, n)^{z'})$. When $z'$ is a constant, this translates to an overall polynomial running time for PA. $\Box$

\smallskip\noindent
{\bf Extending from Top-1 to Top-$k$:}
Our PA algorithm can be modified to return top-$k$ products instead of just the best product. For the tag group $T_r$, once a set of products $S_i$ is built, we compress to form the set $S'_i$. However, every time a cluster representative is selected, instead of deleting all the remaining points in the cluster, we remember $k-1$ products within the cluster and associate them with the cluster representative (and if the cluster has less than $k$ products, we remember and associate all the products with the cluster representative).

When all the $m$ iterations are completed, we can return the top-$k$ products as follows: we first return the best product of $S'_m$ along with the  $k-1$ products associated with it. If the number of associated products are less than $k-1$, the second best cluster representative of $S'_m$ and the set of products associated with it are returned, and so on.

When the approximate top-$k$ products from all tag groups have been returned, the main algorithm returns the overall best top-$k$ products from among them. It can be shown that this approach guarantees an approximation factor for the score of the top-$k$ products returned.

\smallskip\noindent
{\bf Grouping of Tags:} The PA algorithm partitions the set of tags into constant-sized groups. We can employ techniques similar to the grouping of attributes technique for ETT algorithm in order to group related tags together in a principled fashion. However, the bounds and properties of PA algorithm are not affected by this.

\subsection{HC: Hill-Climbing Algorithm}

\begin{algorithm}[!htb]
\caption{ \textbf{HC (Naive Bayes probabilities)}: top-$k$ local optimal products}
\label{alg2}

\begin{algorithmic}[1]
\STATE localOptimaFound $\leftarrow$ false
\STATE Randomly generate a product $\mathnormal{o_s}$(boolean vector) of length $\mathnormal{m}$
\WHILE{localOptimaFound is false}
\FOR{i = 1 to $\mathnormal{m}$}
\STATE $\mathnormal{o_i}$ $\leftarrow$ Neighbors($\mathnormal{o_s}$)
\STATE NeighborScore($\mathnormal{o_s}$) $\leftarrow$ ExactScore($\mathnormal{o_i}$)
\ENDFOR

\IF {Max(NeighborScore($\mathnormal{o_s}$)) $\geq$ ExactScore($\mathnormal{o_s}$)}
\STATE $\mathnormal{o_s}$ $\leftarrow$ $\mathnormal{o_i}$ \COMMENT{$\mathnormal{o_i}$ is highest score in NeighborScore($\mathnormal{o_s}$)}
\ELSE
\STATE localOptima $\leftarrow$ true
\ENDIF

\ENDWHILE
\RETURN $\mathnormal{o_s}$
\end{algorithmic}

\end{algorithm}
Our second approximation algorithm (HC) is based on the generic hill-climbing heuristic, often used for solving complex optimization problems. The algorithm starts from a random solution to the problem (starts at the base of the hill) and then repeatedly improves the solution (walks up the hill) until some condition is maximized (the top of a local hill is reached). In the light of our framework, we generate a random product, i.e., a boolean vector of size $\mathnormal{m}$. At every climbing step, we check all its immediate {\em neighboring} product by examining if a single bit can be flipped to improve the score of the product (by Equation~\ref{eqf}). If there exists such a neighboring product, we proceed to that neighboring product and repeat the climbing step, until a local maximum is reached. Multiple ($k$ or more) random restarts of the hill-climbing technique may lead us to the multiple locally optimal products, of which the top-$k$ may be returned.

Algorithm~\ref{alg2} contains details of our HC algorithm. We illustrate the HC algorithm with an example.

\begin{example}
\textup{
\label{example:hillclimbing}
Consider the boolean dataset in Table~\ref{table:egdatabase} of 8 products, each entry having 4 attributes and 2 tags. We first generate a random product, $1010$ whose score is 1.70. The immediate neighbors of $1010$ are products $0010$, $1110$, $1000$ and $1011$ having scores 1.46, 1.77, 0.89 and 1.76 respectively. Since score of neighbor $1110$ exceeds that of starting product $1010$, we climb to $1110$. The neighbors of $1110$, namely $0110$, $1010$, $1100$ and $1111$ all have scores lesser than that of $1110$ and hence we terminate with $1110$ as the local optimal product (for this example, it also happens to be the global optimum product). $\Box$}
\end{example}
\vspace{-0.10in}

The hill-climbing heuristic is simple, and as we shall discuss later, was found to be extremely effective in our experiments even for large problem instances. However, it has several theoretical limitations. As the following theorem shows, there is no guarantee that it can produce the global optimum; moreover the score of the globally  optimal product may be exponentially  better than the score of a locally optimal product.

\vspace{-0.05in}
\newtheorem{theorem2}{Theorem}
\begin{theorem}
There exists a boolean dataset with $m$ attributes and $z$ tags, and two potential products $o_l$ and $o_g$ where $o_l$ and $o_g$ are a local and the global optimum respectively, such that $\frac{ExactScore(o_g)}{ExactScore(o_l)} =\Omega(n^m)$.
\end{theorem}

{\em Proof}: The exact score of the globally optimum, i.e., the best product (from Equation~\ref{eqf}) can at most be $1$ since $\frac {Pr( \mathnormal{a_i} \mid \mathnormal{T_j}^\prime )}{Pr( \mathnormal{a_i} \mid \mathnormal{T_j})}\  \epsilon\  [\frac{1}{n},\ n]$ and $n$ and $m$ are usually high, so that $\frac{1}{1+n^m} \approx 1$. The local maxima that the HC heuristic may return in the worst case is the product with the lowest score. From Equation~\ref{eqf}, the exact score of the worst case locally optimum product is $\frac{1}{1+n^m} \approx \frac{1}{n^m}$ when $n^m$ is high. Therefore, $\frac{ExactScore(o_g)}{ExactScore(o_l)} \geq \frac{1}{\frac{1}{n^m}} = n^m$.
i.e., $\frac{ExactScore(o_g)}{ExactScore(o_l)} =\Omega(n^m)$

Hence, there exists a dataset for which the expected number of tags of a globally optimum product can be exponentially larger than that of a locally optimum product. $\Box$  

A second limitation of HC is that even for a finding locally optimal product, there is no guarantee on the time taken for termination, and sometimes the convergence may be very slow. This is because certain bits may have to be flipped over and over again, thus forcing the algorithm to explore a huge part of the $2^m$ search space of potential products.


\section{Experiments}
\label{expt}

We conduct a set of comprehensive experiments using both synthetic and real datasets for quantitative and qualitative analysis of our proposed algorithms. Our quantitative performance indicators are (a) {\em efficiency} of the proposed exact and approximation algorithm, and (b) {\em approximation factor} of results produced by the approximation algorithm. The efficiency of our algorithms is measured by the overall execution time and the number of products that are considered from the pool of all possible products, whereas approximation factor is measured as the ratio of the acquired approximate result score to the actual optimal result score. We also conduct a user study through Amazon Mechanical Turk study as well as write interesting case studies to qualitatively assess the results of our algorithms.

\smallskip\noindent
{\bf System configuration$\colon$}
Our prototype system is implemented in Java with JDK 5.0. All experiments were conducted on an Windows XP machine with 3.0Ghz Intel Xeon processor and 2GB RAM. The JVM size is set to 512MB. All numbers are obtained as the average over three runs.

\smallskip\noindent
{\bf Real Camera Dataset$\colon$}
We crawl a real dataset of 100 cameras\footnote{As discussed earlier, the number of products in the dataset is not important for the execution cost; analysis in Figure~\ref{figure:ettproduct}.} listed at Amazon (http://www.amazon.com). The products contain technical details (attributes), besides the tags customers associate with each product. The tags are cleaned by domain experts to remove synonyms, unintelligent and undesirable tags such as {\tt nikon coolpix}, {\tt quali}, {\tt bad}, etc. Since the camera information crawled from Amazon lacks well-defined attributes, we look up Google Products (http://www.google.com/products) to retrieve a rich collection of technical specifications for each product. Each product has 40 boolean attributes, such as {\tt self-timer, face-detection, red-eye fix}, etc; while the tag dictionary includes 40 unique keywords like {\tt lightweight}, {\tt advanced}, {\tt easy}, etc.

\smallskip\noindent
{\bf Real Car Dataset$\colon$}
We crawl another real dataset from Yahoo! Autos (http://autos.yahoo.com/). We focus on new cars listed for the year 2010 spanning 34 different brands. There are several models for each brand, and each model offers several trims.\footnote{\small Trims denote different configurations of standard equipment.}  Since each trim defines a unique attribute-value specification, the total number of trims that we crawl are the 606 products in our dataset.  The products contain technical specifications as well as ratings and reviews, which include pros and cons.  We parse a total of 60 attributes: 25 numeric, and 35 boolean and categorical (which we generalize to boolean) such as  air-conditioning, sunroof, etc . The total number of reviews we extract is 2180. We extract tags from the reviews using the keyword extraction toolkit AlchemyAPI (http://www.alchemyapi.com/api/keyword/). We process the text listed under {\em pros} in each review to identify a set of 15 desirable tags such as \texttt{fuel economy}, \texttt{comfortable interior} and \texttt{stylish exterior}. A car is assigned a tag if one of its reviews contains that keyword.


\smallskip\noindent
{\bf Synthetic Dataset$\colon$}
We generate a large boolean matrix of dimension 10,000 (products)$\times$100 (50 attributes + 50 tags) and randomly choose submatrices of varying sizes, based on our experimental setting. We split the 50 independent and identically distributed  attributes into four groups, where the value is set to 1 with probabilities of 0.75, 0.15, 0.10 and 0.05 respectively. For each of the 50 tags, we pre-define relations by randomly picking a set of attributes that are correlated to it. A tag is set to 1 with probability $p$ if majority of the attributes in its pre-defined relation have boolean 1. For example, assume tag $\mathnormal{T_1}$ is defined to depend on attributes $\mathnormal{A_{13}}$, $\mathnormal{A_{25}}$ and $\mathnormal{A_{40}}$. $\mathnormal{T_1}$ is set to 1 with a probability of 0.67 if 2 out of $\mathnormal{A_{13}}$, $\mathnormal{A_{25}}$ and $\mathnormal{A_{40}}$ are 1.

We use the synthetic datasets for quantitative experiments, while the real data are used in user and case study.

\subsection{Quantitative Results: Performance}

\begin{figure*}[!thb]
\begin{minipage}[b]{0.31\linewidth}
\centering
  \includegraphics[width = \linewidth, height = 3.3cm]{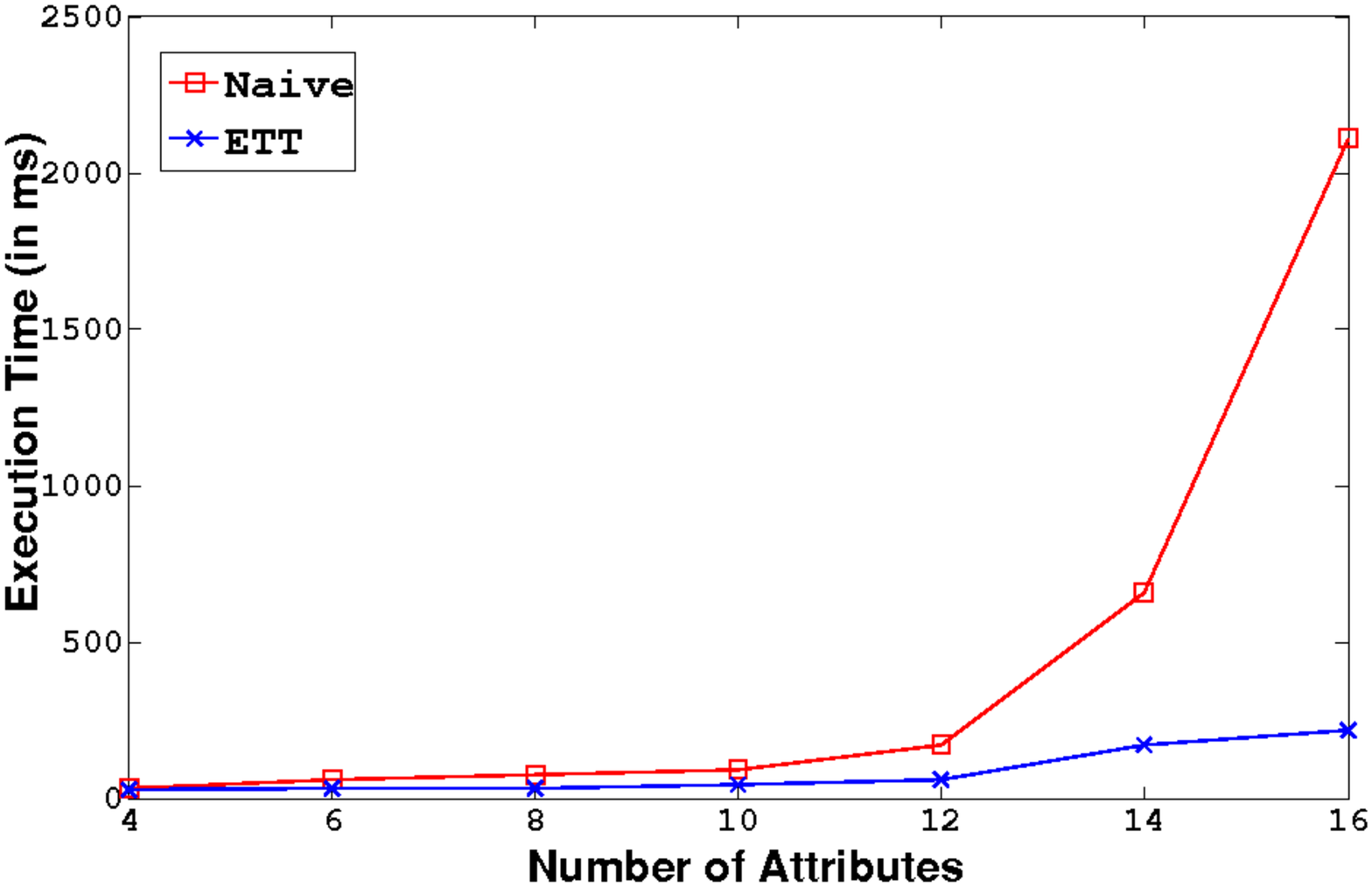}
  \caption{\label{figure:ettnaivetime}\small Execution time for varying $m$ (Synthetic data)}
\end{minipage}
\hspace{0.1cm}
\begin{minipage}[b]{0.31\linewidth}
\centering
  \includegraphics[width = \linewidth, height = 3.3cm]{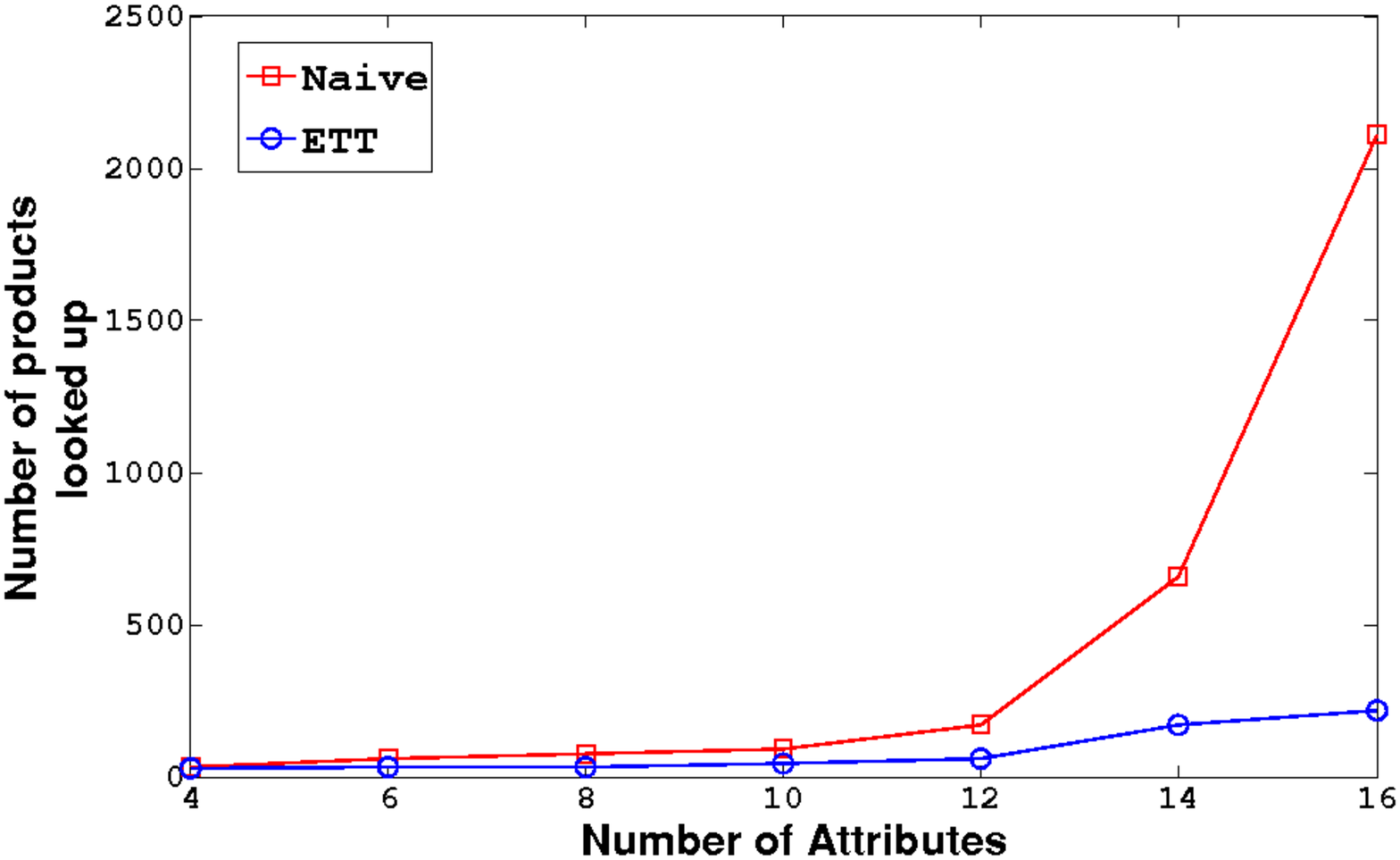}
  \caption{\label{figure:ettnaiveproductbuild}\small Number of products for varying $m$ (Synthetic data)}
\end{minipage}
\hspace{0.1cm}
\begin{minipage}[b]{0.31\linewidth}
\centering
  \includegraphics[width = \linewidth, height = 3.3cm]{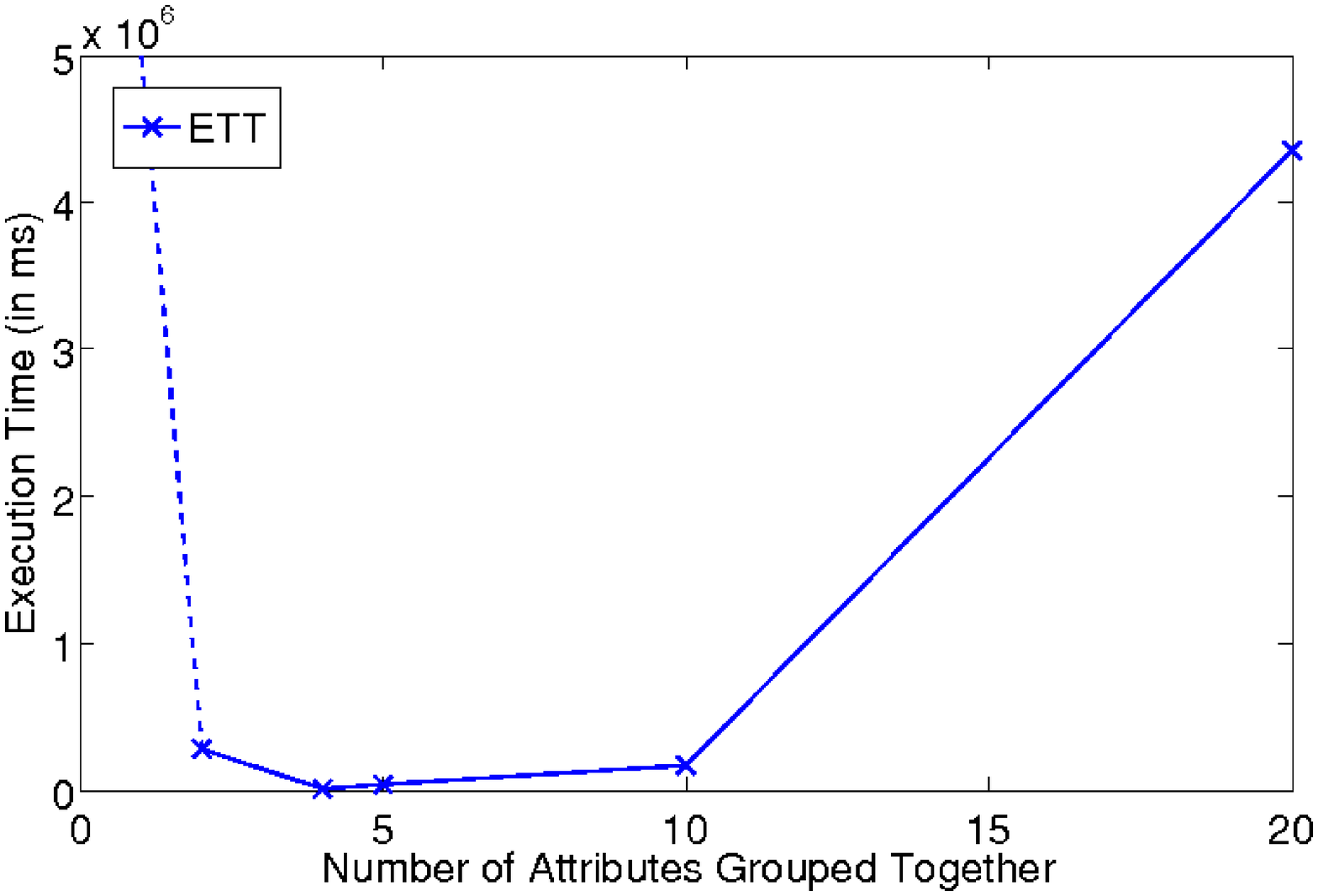}
  \caption{\label{figure:ettgroupingtime}\small Execution time for varying $m^{\prime}$ (Synthetic data)}
\end{minipage}
\hspace{0.1cm}
\begin{minipage}[b]{0.31\linewidth}
\centering
  \includegraphics[width = \linewidth, height = 3.3cm]{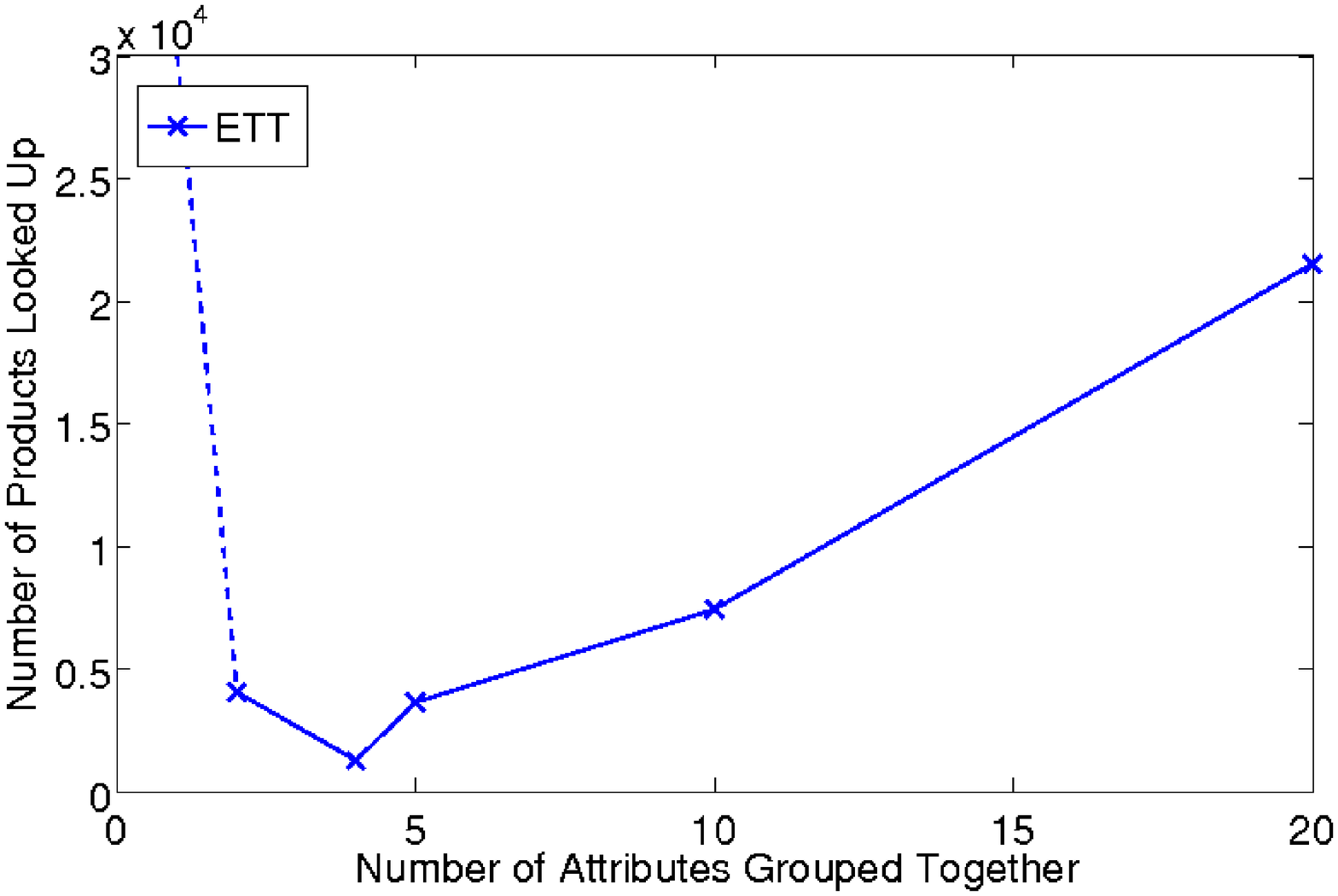}
 \caption{\label{figure:ettgroupingproductbuild}\small Number of products for $m^{\prime}$ (Synthetic data)}
\end{minipage}
\hspace{0.25cm}
\begin{minipage}[b]{0.31\linewidth}
\centering
  \includegraphics[width = \linewidth, height = 3.3cm]{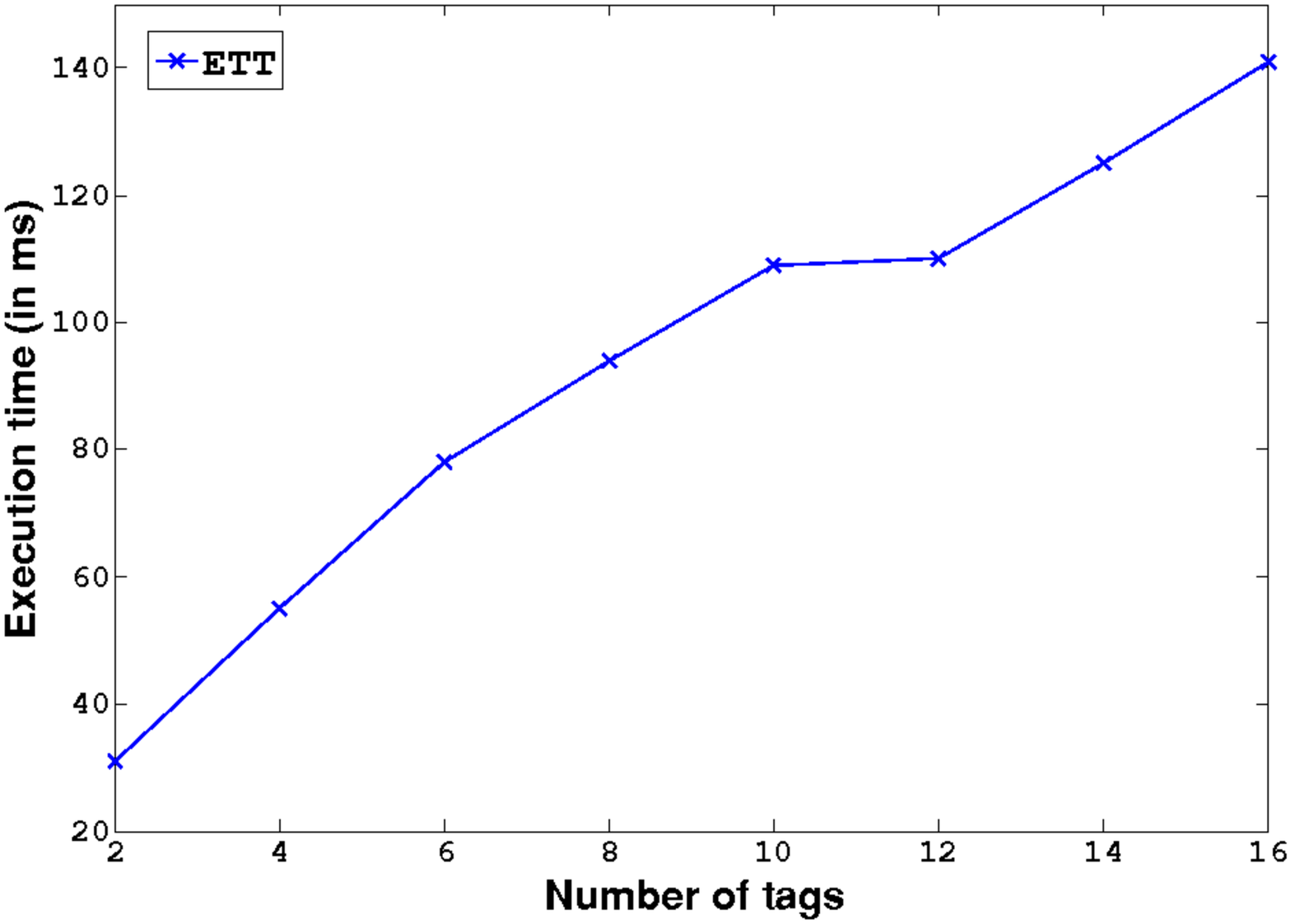}
  \caption{\label{figure:etttag}\small Execution time for varying $z$ (Synthetic data)}
\end{minipage}
\hspace{0.25cm}
\begin{minipage}[b]{0.31\linewidth}
\centering
  \includegraphics[width = \linewidth, height = 3.3cm]{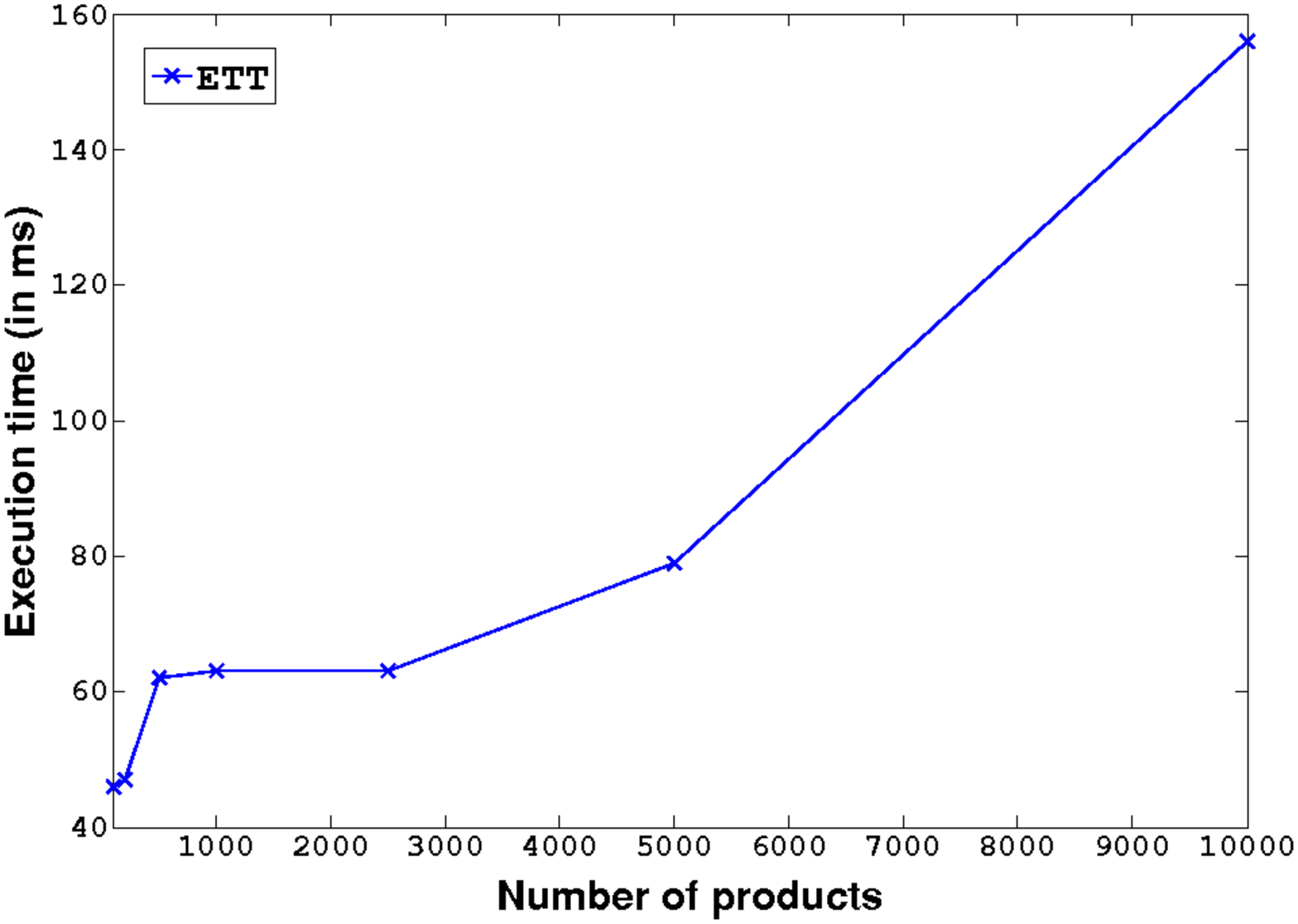}
  \caption{\label{figure:ettproduct}\small Execution time for varying $n$ (Synthetic data)}
\end{minipage}
\hspace{0.1cm}
\end{figure*}

\smallskip\noindent
{\bf Exact Algorithm$\colon$}
We first compare the Naive approach with our ETT. Since the Naive algorithm can only work for small problem instances, we a pick a subset from the synthetic dataset having 1000 products, 16 attributes and 8 tags.
Figures~\ref{figure:ettnaivetime} and~\ref{figure:ettnaiveproductbuild} compare the execution time and the number of candidate products considered, for Naive and ETT respectively, when the number of attributes ($m$) varies (number of products = 1000, number of tags = 8). Note that the number of products considered by ETT is the number of products created in tier-2 by joining products from tier-1. The Naive algorithm considers all $2^m$ products. We used as number of attributes per group $m'=2,2,4,5,4,7,4,6$ for $m= 4,6,8,10,12,14,16,18$ respectively in ETT (more analysis of $m'$ in Figures~\ref{figure:ettgroupingtime} and \ref{figure:ettgroupingproductbuild}). As can be seen, Naive is orders of magnitude slower than ETT.

Next, we study the behavior of attribute groupings on ETT. For a sub-sample picked from our synthetic dataset having 20 attributes, 1000 products and 8 tags, we experiment with different possible attribute groupings,  $m'=$ 1, 2, 4, 5, 10, 20. Figures~\ref{figure:ettgroupingtime} and \ref{figure:ettgroupingproductbuild} shows the effect of $m'$ on the performance of ETT when attributes are grouped arbitrarily.  The execution time and number of products considered for $m'=1$ is not reported in Figures~\ref{figure:ettgroupingtime} and \ref{figure:ettgroupingproductbuild} as it was too slow. The trade-off of choosing $m'$ is: a small $m'$ means there are many short lists in tier-1, so that the cost of joining the lists is high. In contrast, a large $m'$ indicates fewer but longer lists in tier-1 resulting in increased cost of creating the lists. We observe that the best balance is struck when $m'=4$ attributes forming 5 lists, each having $2^4$=16 products. 

We also employed the grouping of attributes technique in Section~\ref{optalgo1} to partition the set of 20 attributes, and investigate if the execution time and number of products considered improves (i.e., decreases). We create a graph of 20 nodes (corresponding to the 20 attributes) and $^{20}C_{2}$ = 190 edges. We use the absolute value of the Pearson correlation to determine the edge weight because even if two attributes are anti-correlated, they should be grouped together and the $1$ from one will be combined with the $0$ from the other to create a high-scored entry (we use 0.5 for don't care). Then, we employ a graph partitioning algorithm for partitioning the 20 attributes into as many groups as the desired number of lists. Specifically, we use the publicly available software METIS-4.0 (http://glaros.dtc.umn.edu/gkhome/views/metis) for partitioning the attributes. We observe that when we partition the 20 attributes into 4 clusters (i.e., $m'=5$ attributes forming 4 lists), the execution time is $\frac{1}{6}$ the execution time in case of arbitrary grouping of attributes (Figure~\ref{figure:ettgroupingtime}); the number of products looked up by ETT decreases from 3639 to 1713 (Note that the number of products looked up by Naive for this data is 1048576). Again, if we partition the 20 attributes into 5 clusters (i.e., $m'=4$ attributes forming 5 lists), the execution time and the number of products looked up by ETT remains the same as in case of arbitrary grouping of attributes (Figure~\ref{figure:ettgroupingtime}). This is because clustering of attributes into 4 groups generated stronger partitioning (i.e., attributes grouped together have stronger correlation) than clustering of attributes into 5 groups. Therefore, if the data is highly correlated and yields well-defined clusters or partitions, ETT benefits significantly by employing principled grouping of attributes.       

Next, we vary the number of tags $z$ and number of products $n$ in the dataset to study the behavior of ETT. We pick a subset from the synthetic dataset having 1000 products, 16 attributes and 16 tags, and consider further subsets of this dataset. Figure~\ref{figure:etttag} reflects the change in execution time with increasing number of tags for the synthetic data (number of products = 1000, number of attributes = 12, attribute grouping = 3). The increase in number of tags increases the number of GetNext() operations in ETT, and hence the running time rises steadily. Figure~\ref{figure:ettproduct} depicts how an increase in the number of products in the dataset (number of attributes = 12, number of tags = 8, attribute grouping = 3) barely affects the running time of ETT since an initialization step calculates all conditional tag-attribute probabilities.

{\bf Approximation Algorithms$\colon$}
We observe in Figure~\ref{figure:ettnaivetime} that the execution time of ETT outperforms that of Naive, for moderate data instances. However, ETT is extremely slow beyond number of attributes ($m$) = 16, which makes it unsuitable for large real-world datasets having many attributes and tags. Therefore, we move to our approximation algorithms HC and PA, and compare their execution time and sub-optimal product score in Table~\ref{table:appcomparison}.  We pick three different subsets of 1000 products from the synthetic dataset: (number of attributes = 8, number of tags = 4), (number of attributes = 12, number of tags = 8) and (number of attributes = 16, number of tags = 8). We execute PA algorithm at an approximation factor (0.8) and HC algorithm without multiple random-restart. The execution time of PA for a moderately large dataset (1000 products, 16 attributes and 12 tags) indicates that it is unlikely to scale to large (real) datasets. Nevertheless, it is the only algorithm, of the two, which provides worst case guarantees in both time complexity and result quality. On the other hand, our HC algorithm is quite effective even as the number of attributes and tags increases. For the dataset (1000 products, 16 attributes and 12 tags), HC is $10^4$ times faster than PA, while the quality of sub-optimal product remains comparable. 
Table~\ref{table:appcomparison} shows a situation (n=1000, m=8, z=4) when HC takes similar amount of time as PA to retrieve identical sub-optimal product and another situation (n=1000, m=12, z=8) when HC takes lesser amount of time than PA to retrieve a sub-optimal product inferior in quality to that returned by PA. However, multiple random restarts of HC may lead us to better sub-optimal product(s). 

\vspace{-0.05in}
\begin{table}[!htb]
\centering
\tbl{Comparison: Approximation Algorithms}{
\begin{tabular*}{\linewidth}{|p{2.9cm}|p{2.9cm}|p{2.0cm}|p{2.0cm}|p{2.0cm}|}
\hline
& {\begin{minipage}[b]{\linewidth}
     \centering
     {\small Approx-\\imation\\Algorithms}
 \end{minipage}
 } & {\begin{minipage}[b]{\linewidth}
     \centering
     {\small n=1000\\m=8\\z=4}
 \end{minipage}} & {\begin{minipage}[b]{\linewidth}
     \centering
     {\small n=1000\\m=12\\z=8}
 \end{minipage}} & {\begin{minipage}[b]{\linewidth}
     \centering
     {\small n=1000\\m=16\\z=12}
 \end{minipage}} \\ \hline
\multirow{2}{*}{\begin{minipage}[t]{\linewidth}
                   \centering
                   {\small  Execution\\Time(in ms)}
               \end{minipage}} & \centering HC & \centering 47.0 & \centering 55.0
& \begin{minipage}[t]{\linewidth}\centering {61.0} \end{minipage} \\
\cline{2-5}
& \centering PA & \centering 62.0 & \centering 1235.0 
&\begin{minipage}[t]{\linewidth}\centering {710032.0} \end{minipage}\\
\hline
\multirow{2}{*}{\begin{minipage}[t]{\linewidth}
                   \centering
                   {\small  Sub-optimal Product Score}
               \end{minipage}} & \centering HC & \centering 3.077 & \centering
3.596 & \begin{minipage}[t]{\linewidth}\centering {9.034}
\end{minipage}\\ \cline{2-5}
& \centering PA & \centering 3.077 & \centering 5.527 &
\begin{minipage}[t]{\linewidth}\centering {9.034} \end{minipage} \\
\hline
\end{tabular*}}
\label{table:appcomparison}
\end{table}

\subsection{Qualitative Results: User Study}
We now validate how designers can leverage existing product information to design new products catering different groups of people in a user study conducted on Amazon Mechanical Turk (https://www.mturk.com) on the real camera dataset. We also consult DPreview (http://www.dpreview.com), a website about digital cameras and digital photography. There are two parts to our user study. Each part of the study involves thirty independent single-user tasks. Each task is conducted in two phases: {\em User Knowledge Phase} where we estimate the users' background and {\em User Judgment Phase} where we collect users' responses to our questions.

In the first part of our study, we build four new cameras (two digital compact and two digital slr) using our HC algorithm 
by considering tag sets corresponding to compact cameras and slr cameras respectively. We present these four new cameras along with four existing popular cameras (presented anonymously) and observe that 65\% of users choose the new cameras, over the existing ones. For example, users overwhelmingly prefer our new compact digital camera over Nikon Coolpix L22 because the former supports both automatic and manual focus while the latter does not, thus validating how our techniques can benefit designers.

\begin{figure}[!hbt]
\centering
\includegraphics[width=3.1in,height=2.0in]{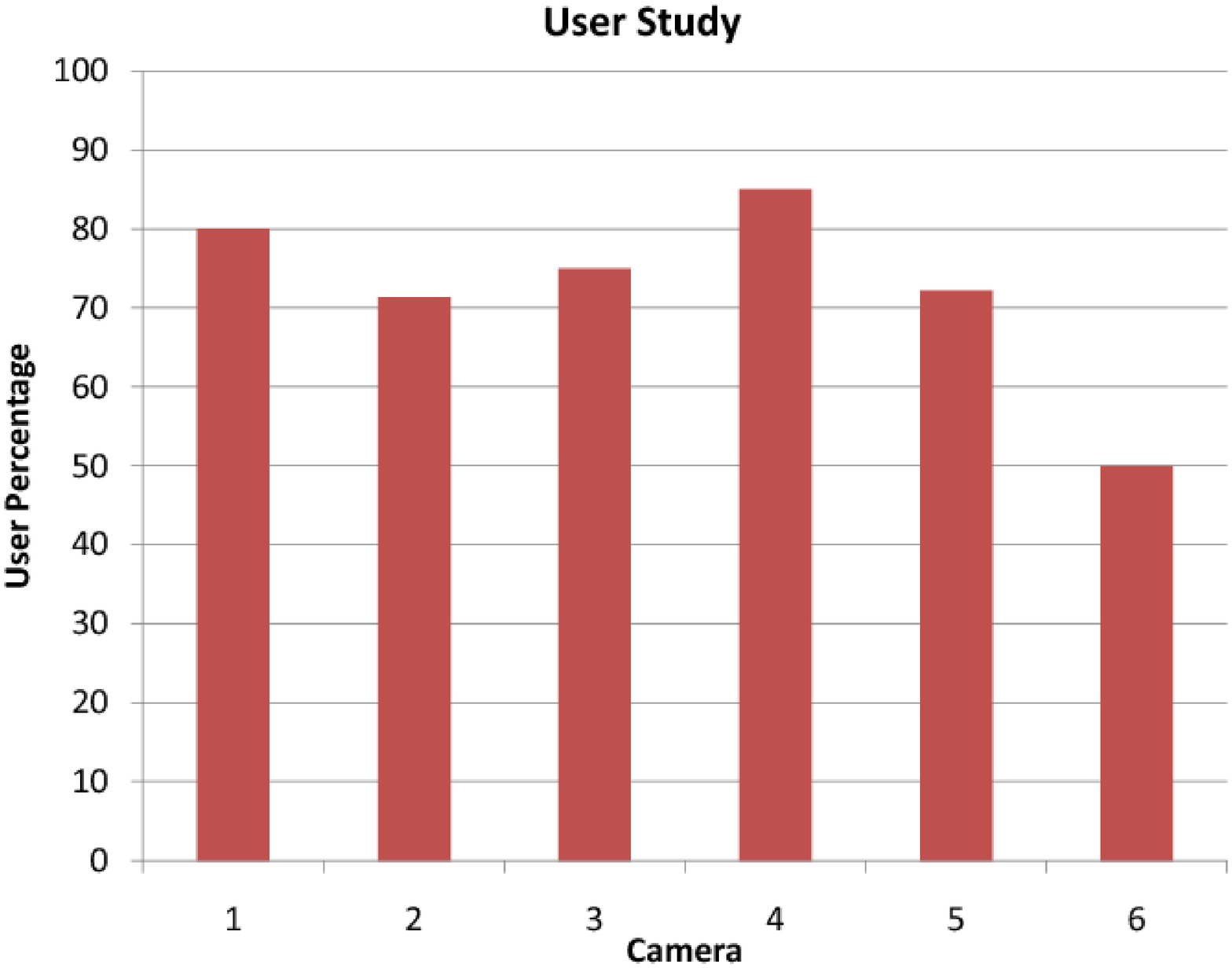}
\vspace{-0.05in}
\caption{Users Classify Cameras Correctly}
\label{figure:userstudy}
\end{figure}
 
The second part of the study concerns six new cameras designed for three groups of people$\colon$ young students, old retired and professional photographers. Domain experts identify and label three overlapping sets of tags from the camera dataset's complete tag vocabulary, one set for each group and we then build two potential new cameras for each of the three groups. For each of the six new cameras thus built, we ask users to assign at least five tags by looking up the complete camera tag vocabulary, provided to them. We observe that majority of the users rightly classify the six cameras into the three groups. The correctness of the classification is validated by comparing the tags received for a camera to the three tag sets identified by domain experts; we also validate the correctness by consulting data available in Dpreview. For example, the cameras designed by leveraging tags corresponding to professional photographers draw tags like {\tt advanced}, {\tt high iso}, etc. while cameras designed by leveraging tags corresponding to old retired draw tags like {\tt lightweight}, {\tt easy}, etc.  Figure~\ref{figure:userstudy} shows the percentage of users classifying the six cameras correctly. Thus, our technique can help designers build new products that are likely to attract desirable tags from different groups of people.

\subsection{Qualitative Results: Case Study}
We present few interesting anecdotal results returned by our framework on the real car dataset to validate that our algorithms help us draw interesting conclusions about the desirability of certain car specifications (attribute values). Our HC algorithm indicates that cars having child safety door locks, 4-wheel anti lock brakes, AM/FM radio, keyless entry, telescoping steering wheel, compass, air filter, trunk light, smart coinholder and cup holder, etc. are the features likely to elicit positive feedback from the customers (i.e., features that maximize the set of desirable tags for real car dataset). When we design new cars by considering only those car instances as our training set which have received the tag {\tt economy}, we observe that some luxury features like heated seats, in-dash CD changer system, sunroof/moonroof and leather upholstery are returned by our framework. This indicates that users prefer selective luxury features when buying economy cars. Also, sports cars designed using our algorithm (by considering only those car instances as our training set which have received the tag {\tt sports}) are found to contain safety features, thereby indicating that safety features have become a high priority requirement for users buying sports cars.

\section{Related Work}
\label{rel}

\smallskip\noindent
{\bf Tag prediction$\colon$}
The dynamics of social tagging has been an active research area in recent years, with several papers focusing on the tag prediction problem. A recent work~\cite{Yin10} proposes a probabilistic model for personalized tag prediction and employs the Naive Bayes classifier. Related research in text mining~\cite{twitter2010} found that the Naive Bayes classifier performs better than SVM and CRF in classifying blog sentiments. Another study that indirectly supports the use of Naive Bayes for tag prediction is done by Heymann et al.~\cite{socialtag2008}, who found that tag-based association rules can produce very high-precision predictions. The process of collaborative tagging has been studied in \cite{usagepattern2006} and~\cite{DBLP:journals/corr/abs-cs-0508082}; the challenges associated with tag recommendations for collaborative tagging systems have been discussed in~\cite{jaeschke2012challenges}.~\cite{KimH2010etal} develops a new unique recommendation algorithm via collaborative tags of users to provide enhanced recommendation quality and to overcome some of the limitations in collaborative filtering systems. Other related work investigates tag suggestion, usually from a collaborative filtering and UI perspective; for example with 
images~\cite{DBLP:journals/pami/WuJJ13} and blog posts~\cite{Mishne2006}. Due to the high popularity of social bookmarking systems,~\cite{Michlmayr07learninguser} proposes a technique for building a user
profile from a user's tagging behaviour, thereby indicating the usefulness of tags in representing user opinion.  

\smallskip\noindent
{\bf Item design$\colon$}
The problem of item design has been studied by many disciplines including economics, industrial engineering and computer science~\cite{Selkar00}. Optimal item design or positioning is a well studied problem in Operations Research and Marketing. Shocker \textit{et al.}~\cite{shocker1974} first represented products and consumer preferences as points in a joint attribute space. Later, several techniques \cite{Albers80}, \cite{Albritton} were developed to design/position a new item. Work in this domain requires direct involvement of consumers, who choose preferences from a set of existing alternative products.~\cite{manufacturing06} reviews the role of data mining in manufacturing engineering, in particular production processes, operations, fault detection, maintenance, decision support, and product quality improvement. Tucker proposes a machine learning model to capture emerging customer preference trends within the market space in~\cite{tucker2011}. Miah et al.~\cite{maxvisibility2009} study the problem of selecting product snippets given a user query log, in order for the designed snippet to be returned by the maximum number of queries.  Our problem is different because the tags of a product are correlated to its attributes (through a classifier), whereas the queries in ~\cite{maxvisibility2009} are boolean section conditions.
However, none of these works has studied the problem of item design in relation to social collaborative tagging.

\smallskip\noindent
{\bf Top-k algorithms$\colon$}
Our top-$k$ pipelined algorithm is inspired by the rich work on top-$k$ algorithms \cite{topk2001}, \cite{rankjoin2009}. A recent survey by Ilyas et al.~\cite{topksurvey2008} covers many of the important results in this area. The classic setting of these works is that each list contains an attribute of an object and a monotone aggregate function is used for ranking. The top tier of our pipelined top-$k$ algorithm is adapted from this setting, where each list has the probability of a tag for each assignment of attribute values. In contrast, in the bottom tier of our algorithm, an entry from one list can match with any entry from the other lists. This setting is adapted from the problem of top-k join~\cite{rankjoin2009}.

\section{Conclusions}
\label{conc}
In this paper we consider the novel problem of leveraging online collaborative tagging in product design. We formally define the Tag Maximization problem, investigate its computational complexity, and propose several principled algorithms that are shown to work well in practice.
Our work is a preliminary look at a very novel area of research, and there appear to be many exciting directions of future research. Our immediate focus is to extend our work to include tag prediction using other classifiers, such as decision trees, SVMs, and regression trees (the latter is applicable when we wish to predict the frequency of occurrence of desirable tags attracted by products). We also intend to evaluate the applicability of our proposed framework to other novel applications, e.g., guide recommender systems recommend better vacation travel itineraries by tracking tag history, help online authors write better blogs, and others.
\vspace{-0.05in}

\bibliographystyle{acmsmall}
\bibliography{acmsmall-sample}


\end{document}